%% file: paper.tex
\documentclass[sigconf]{acmart}

\acmConference[SEAMS 2022]{The 17th Symposium on Software Engineering for Adaptive and Self-Managing Systems}{May 21–29, 2022}{Pittsburgh, PA, USA}

\usepackage{amsmath,amsfonts}
\usepackage{esvect}
\usepackage{graphicx}
\usepackage{subfigure}
\usepackage{booktabs}
\usepackage{tabularx}
\usepackage{multirow}
\usepackage{balance}
\usepackage{xcolor}
\usepackage{enumitem}
\usepackage{mathtools}
\usepackage{wrapfig}
\usepackage{ragged2e}
\usepackage[all]{nowidow}
\usepackage{stfloats}
\usepackage{anyfontsize}
\usepackage{algorithm}

\usepackage{color}
\usepackage{colortbl}

\newcolumntype{Y}{>{\centering\arraybackslash}X}

\usepackage{colortbl}
\usepackage{listings}

\theoremstyle{definition}

\newtheorem{dfn}{Definition}[section]

\definecolor{javared}{rgb}{0.6,0,0} 
\definecolor{javagreen}{rgb}{0.25,0.5,0.35} 
\definecolor{javapurple}{rgb}{0.5,0,0.35} 
\definecolor{javadocblue}{rgb}{0.25,0.35,0.75} 
\definecolor{javagrey}{rgb}{0.46,0.45,0.48} 

\newcommand\resq[1]{
\noindent 
\fcolorbox{green!40!black}{green!5}{\noindent 
 \parbox{0.98\columnwidth}{\noindent  #1}}\\
}

\lstdefinestyle{Alg}{
  basicstyle=\ttfamily\footnotesize,
  breaklines=true,
  tabsize=2,
  mathescape,
  numbers=left,
  xleftmargin=2.5em,
  xrightmargin=0.5em,
  frame=tb,
  framexleftmargin=2em,
  emph={Algorithm,Input,Output,for,each,do,if,else,Function,while,let,be,repeat,until,return,times,and,or,break,in,then,},
  emphstyle={\textbf},
  escapechar=?,
  morecomment=[l][\color{javagreen}]{//},
  columns=flexible,
}

\usepackage{framed}
\usepackage{tikz}
\definecolor{light-gray}{gray}{0.9}
\usepackage[framemethod=TikZ]{mdframed}
\mdfsetup{skipabove=5pt,skipbelow=3pt}
\usepackage{lipsum}
\mdfdefinestyle{RQFrame}{%
	linecolor=black,
	outerlinewidth=0.15pt,
	roundcorner=3pt,
	innertopmargin=2pt,
	innerbottommargin=2pt,
	innerrightmargin=4pt,
	innerleftmargin=4pt,
	backgroundcolor=light-gray}

\copyrightyear{2022}
\acmYear{2022}
\setcopyright{acmcopyright}\acmConference[SEAMS '22]{17th International Symposium on Software Engineering for Adaptive and Self-Managing Systems}{May 18--23, 2022}{PITTSBURGH, PA, USA}
\acmBooktitle{17th International Symposium on Software Engineering for Adaptive and Self-Managing Systems (SEAMS '22), May 18--23, 2022, PITTSBURGH, PA, USA}
\acmPrice{15.00}
\acmDOI{10.1145/3524844.3528053}
\acmISBN{978-1-4503-9305-8/22/05}

\begin{document}

\title[Learning Self-adaptations for IoT Networks: A Genetic Programming Approach]{Learning Self-adaptations for IoT Networks:\\ A Genetic Programming Approach}

\author{Jia Li, Shiva Nejati and  Mehrdad Sabetzadeh}
\affiliation{%
  \institution{University of Ottawa}
  \country{Canada}
}
\email{{jli714,snejati,m.sabetzadeh}@uottawa.ca}


\input{abstract.tex}

\maketitle

\input{intro.tex}
\input{motivation.tex}

\input{approach.tex}

\input{algorithmGP.tex}
\input{eval.tex}

\input{results.tex}
\input{Rq2.tex}

\input{threats.tex}

\input{related.tex}
\input{conclusions.tex}

\begin{acks}
We gratefully acknowledge funding from  NSERC of Canada under the Discovery and Discovery Accelerator programs. We thank Seung Yeob Shin for discussions at the early stages of this work. This research was enabled in part by support provided by Compute Ontario and WestGrid (https://www.westgrid.ca/) and Compute Canada (www.computecanada.ca).
\end{acks}

\bibliographystyle{ACM-Reference-Format}
\balance
\bibliography{paper.bib}

\end{document}

%% file: abstract.tex

\begin{abstract}
Internet of Things (IoT) is a pivotal technology in application domains that require connectivity and interoperability between large numbers of devices.  IoT systems predominantly use a software-defined network (SDN) architecture as their core communication backbone. This architecture offers several advantages, including the flexibility to make IoT networks self-adaptive through software programmability. In general, self-adaptation solutions need to periodically monitor, reason about, and adapt a running system. The adaptation step involves generating an adaptation strategy and applying it to the running system whenever an anomaly arises. In this paper, we argue that, rather than generating individual adaptation strategies, the goal should be to adapt the logic / code of the running system in such a way that the system itself would learn how to steer clear of future anomalies, without triggering self-adaptation too frequently. We instantiate and empirically assess this idea in the context of IoT networks. Specifically, using genetic programming (GP), we propose a self-adaptation solution that continuously learns and updates the control constructs in the data-forwarding logic of SDN-based IoT networks. Our evaluation, performed using open-source synthetic and industrial data, indicates that, compared to a baseline adaptation technique that attempts to generate individual adaptations, our GP-based approach is more effective in resolving network congestion, and further, reduces the frequency of adaptation interventions over time. In addition, we compare our approach against a standard  data-forwarding algorithm from the network literature, demonstrating that our approach \hbox{significantly reduces packet loss.}
\end{abstract}

%



%% file: intro.tex
\vspace*{-.05cm}
\section{Introduction}
\label{sec:intro}
A major challenge when engineering complex systems is to ensure that these systems  meet their quality-of-service criteria and are reliable in the face of uncertainty. Self-adaptation is a promising approach for addressing this challenge. 
The idea behind self-adaptation is that engineers take an existing system, specify its expected qualities and objectives as well as strategies to achieve these objectives, and build capabilities into the system in a way that enables the system to adjust itself to changes during operation~\cite{DBLP:books/daglib/p/GarlanSC09}.

Many software-intensive systems can benefit from self-adaptivity. A particularly pertinent domain where self-adaptation is useful is Internet of Things (IoT). IoT envisions highly complex systems composed of a large number of smart devices that are implanted with sensors and actuators, and interconnected through the Internet or other network technology~\cite{IoT17}. There are various sources of uncertainty in IoT applications. These include, among others, the dynamic and rapidly changing environment in which IoT applications operate and the scarcity and limitations of resources, e.g., network bandwidth, battery power, and data-storage capacity. In addition, since IoT rollouts are typically very expensive, IoT applications are expected  to live for a long period of time and work reliably throughout~\cite{Weyns:18,Iftikhar:17}. These factors necessitate some degree of self-adaptivity in IoT systems. Notably, given their often limited bandwidth, IoT networks are prone to \emph{congestion} when there is a burst in demand. For instance, in an IoT-based emergency management system, such bursts can occur when a disaster situation, e.g., a flood, is unfolding. Our ultimate goal in this paper is to develop a self-adaptive approach for resolving congestion in IoT networks.


Self-adaptation has been studied for many years~\cite{Moreno:15,PaucarB19,JahanRWGPMC20,ChengRM13,DBLP:journals/taas/SalehieT09}. The existing self-adaptation approaches include  model-based~\cite{DBLP:conf/models/DeVriesC17,Ramirez:10}, control-based~\cite{Filieri:15}, requirements-based~\cite{DBLP:conf/icse/AlrajehCL20}, and more recently, learning-based~\cite{DBLP:conf/seams/GheibiWQ21} solutions. At the heart of all self-adaptation solutions, there is a planning step that generates or determines an adaptation strategy to adapt the running system once an anomaly is detected~\cite{DBLP:books/daglib/p/GarlanSC09}. Adaptation strategies may be composed of fixed and pre-defined actions identified based on domain knowledge, or they may be new behaviours or entities introduced at run-time~\cite{DBLP:journals/taas/SalehieT09}. Irrespective of the type of adaptation, the knowledge about how to generate adaptation strategies is often concentrated in the planner: The running system is merely modified by the planner to be able to handle an anomaly as seen in a specific time and context; the running system is not necessarily improved in a way that it can better respond to future anomalies on its own. As we illustrate in our motivating example of Section~\ref{sec:motivate}, having to repeatedly invoke the planner for each adaptation can be both expensive and ineffective. 
The main idea that we put forward in this paper is as follows: \emph{The self-adaptation planner should attempt to improve the logic / code of the running system such that the system itself would learn how to steer clear of future anomalies,  without triggering self-adaptation too frequently.} The need for self-adaptation is never entirely eliminated, especially noting that the environment changes over time. Nonetheless, a system augmented with such a self-adaptation planner is likely to be more efficient, more robust to changes in the environment (e.g., varying network requests), and less in \hbox{need of adaptation intervention.}

Furthermore, existing self-adaptation research focuses on modifying a running system via producing individual and concrete elements, e.g., configuration values~\cite{Iftikhar:17}. In contrast, we take a  \emph{generative} approach~\cite{FeldtY20}, modifying the logic of the running system which in turn generates the concrete elements.

We instantiate the above-described vision of self-adaptation for addressing congestion control in \emph{software-defined networks (SDNs)}. IoT systems are predominantly built and deployed over SDNs due to the flexibility offered by this networking architecture~\cite{Lopes:16}. In an SDN,  network control is transferred from local fixed-behaviour controllers distributed over a set of switches to a centralized and programmable software controller~\cite{SDN:15}. More specifically, we address the self-adaptation of SDN data-forwarding algorithms in order to resolve network congestion in real time. Our self-adaptation technique employs \emph{genetic programming (GP)}~\cite{koza1992genetic,poli2008field} to enhance the  programmable controller of an SDN with the well-known MAPE-K self-adaptation  loop~\cite{kephart:03,Moreno:15}. Our approach (i)~periodically monitors the SDN to check if it is congested and generates a model of the SDN  to be used for congestion resolution; (ii)~applies GP to evolve the logic of the SDN data-forwarding algorithm such that congestion is resolved and, further, the transmission delay and the changes introduced in the existing  data-transmission routes  are minimized; and then (iii)~modifies  existing transmission routes to resolve the current congestion, and updates the logic of the SDN data-forwarding algorithm to mitigate future congestion. Exploiting the global network view provided by an SDN for modifying the controller and resolving congestion in real time is not new. However, existing approaches  adapt  the network management logic/code using  \emph{pre-defined} rules~\cite{Yashar:12}. In contrast, our approach uses GP to modify the data-forwarding control constructs in an evolving manner and without reliance \hbox{on any fixed rules.}

We evaluate our generative, adaptive approach on $18$ synthetic and one industrial IoT network. The industrial network, which is the backbone of an IoT-based emergency management system~\cite{ShinNSB0Z20}, is prone to congestion when the volume of demand increases during emergencies. We compare our approach with two baselines: (1) a self-adaptive technique from the SEAMS community that attempts to resolve congestion by generating individual adaptations (i.e., individual data-transmission routes) without optimizing the SDN routing algorithms~\cite{ShinNSB0Z20}; and (2)~a standard data-forwarding algorithm from the network community that uses pre-defined rules (heuristics) to optimize SDN control at runtime and resolve congestion~\cite{Coltun:08,Cisco:05}. Our results indicate that our approach successfully resolves congestion in all the 18 networks considered while the adaptive, but non-generative baseline fails to resolve congestion in four of the networks with probabilities ranging from $10$\% to $66$\%. In addition, compared to this baseline, our approach reduces the average number of congestion occurrences, and hence, the number of adaptation rounds necessary. Compared to the baseline from the network community, our approach is able to significantly reduce packet loss for the industrial network. Finally, we empirically demonstrate that the execution time of our approach is  linear in network size and the amount of data traffic over time, making our approach suitable for online adaptation. 

\emph{Structure.} Section~\ref{sec:motivate} motivates the paper. Section~\ref{sec:selfcontrol} presents our approach. Section~\ref{sec:eval} describes our evaluation. Section~\ref{sec:related} compares with related work. Section~\ref{sec:conclusions} concludes the paper. 

%% file: motivation.tex
\vspace*{-.3em}
\section{Motivating Example}
\label{sec:motivate}
We motivate our approach through an example. Network-traffic management is about assigning flow paths to network requests such that the entire network is optimally utilized~\cite{Noormohammadpour18,Alizadeh:10,Agarwal:13,Betzler:16}. A flow path is a directed path of network links.  Many network systems use a standard, lightweight data-forwarding algorithm, known as Open Shortest Path First (OSPF)~\cite{Coltun:08,Cisco:05}. OSPF generates a flow path for a data-transmission request by computing the shortest weighted path between the source and destination nodes of the request.  

\begin{figure}[t]
	\centerline{\includegraphics[width=\columnwidth]{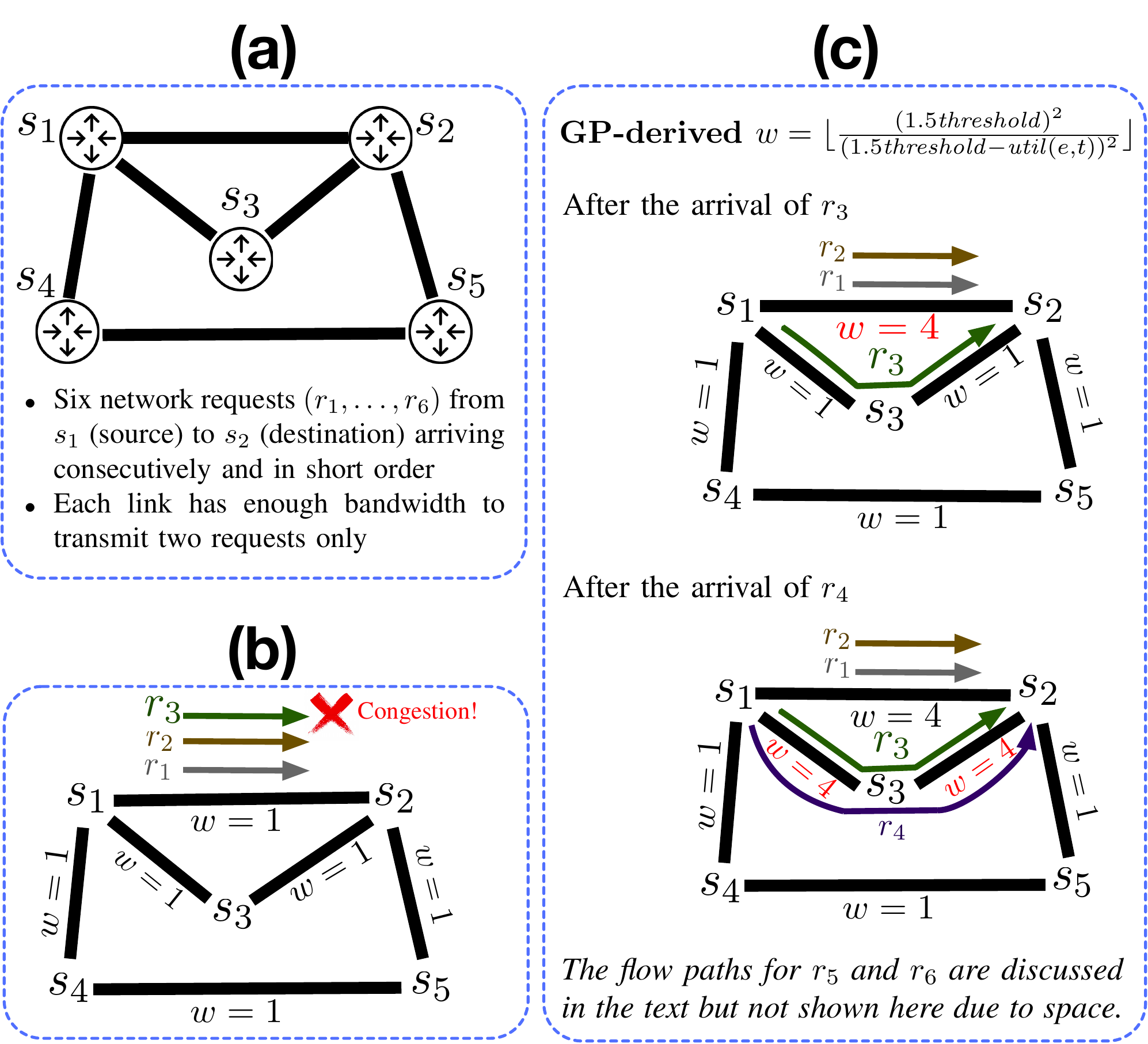}}
	\vspace{-0.7em}
	\caption{(a) A simple network alongside the description of network requests; (b) Output of a baseline data-forwarding algorithm; and (c) Output of data-forwarding after adaptation by our approach using genetic programming.}
	\label{fig:example}
	\vspace{-1.5em}
\end{figure}

Figure~\ref{fig:example}(a) shows a network with five nodes (switches) and six links. Suppose that there are six  requests, $r_1$, \ldots, $r_6$ from source node $s_1$ to destination node $s_2$, and that they arrive in sequence (first $r_1$, then $r_2$, \ldots) with a few seconds in between. Further, all link weights are set to one. OSPF creates the flow path $s_1 \rightarrow s_2$ for every request.
Assume that each flow utilizes $30$\% of the bandwidth of each link. Further, assume that we have a threshold of $80$\% for link utilization, above which we consider a link to be congested. 
Each link in our example then has enough bandwidth to transmit two flows before it is considered congested. Consequently, link $s_1 \rightarrow s_2$ will be congested after the arrival of $r_3$; see Figure~\ref{fig:example}(b). This leads to an invocation of
the self-adaptation planning step to resolve the congestion. A common approach for congestion resolution is through combinatorial optimization, where some flow paths are re-routed such that the data stream passing through each link remains below the utilization threshold~\cite{FortzT02,DBLP:conf/cscwd/LiuZGQ18,6980429}. This optimization can  be done using various methods, e.g., graph optimization~\cite{FortzT02}, mixed-integer linear programming~\cite{Agarwal:13}, \hbox{local search~\cite{FortzT00,FortzT04}, and genetic algorithms~\cite{ShinNSB0Z20}.}

Optimization by re-routing flow paths has a major drawback. While optimizing at the level of individual flow paths may solve the currently observed  congestion, doing so does not improve the logic of OSPF and thus does not contribute to congestion avoidance in the future. For instance, in Figure~\ref{fig:example}(b), the arrival of $r_3$ will cause congestion. Re-routing $r_3$ over $s_1 \rightarrow s_3 \rightarrow s_2$ will resolve this congestion. But, upon the arrival of $r_4$, OSPF will yet again select the flow path $s_1 \rightarrow s_2$ for $r_4$ (since this is the shortest path between $s_1$ and $s_2$) just to find the $s_1 \rightarrow s_2$ link congested again. This in turn necessitates the self-adaptation planning step to be invoked for $r_4$ as well. In a similar vein, the arrival of $r_5$ and $r_6$ will cause congestion, prompting further calls to self-adaptation planning.

Our proposal is that, instead of optimizing flow paths, one should optimize the logic of OSPF whenever a congestion is detected. In this paper, we optimize and update the link-weight formula that OSPF uses. The hypothesis here is that an optimized link-weight formula not only can resolve the existing congestion, but can simultaneously also make OSPF more intelligent towards congestion avoidance in the future, thus reducing the number of times that the self-adaptation planning step has to be invoked. 
Below, we illustrate our approach using the example of Figure~\ref{fig:example}. 

We use genetic programming (GP) for self-adaptation planning, whereby we dynamically learn link-weight functions that not only help resolve an existing congestion but also help steer clear of future ones.
Initially, and for requests $r_1$ and $r_2$, our approach does exactly as the standard OSPF would do, since there is no congestion. Upon the arrival of $r_3$ and the detection of congestion, i.e., the situation in Figure~\ref{fig:example}(b), our self-adaptation approach kicks in and automatically computes a link-weight formula such as the following: $ \lfloor \frac{(1.5\mathit{threshold)^2}}{(1.5\mathit{threshold} - \mathit{util}(e, t))^2}\rfloor$. 
In this formula,  $\mathit{util}(e, t)$ is the utilization percentage of link $e$ at time $t$, and  $\mathit{threshold}$ is a constant parameter describing the utilization threshold above which the network is considered to be congested. In our example, $\mathit{threshold}=0.8$, and for each link, $\mathit{util}(e, t)$ is 0.3 if one flow passes through $e$, and is 0.6 if two flows pass through $e$.  Given that  $r_1$ and $r_2$ are already routed through $s_1 \rightarrow s_2$, the weight of \hbox{$s_1 \rightarrow s_2$} computed by this formula becomes $w=4$; see Figure~\ref{fig:example}(c) after $r_3$'s arrival. As a result, OSPF selects path $s_1 \rightarrow s_3 \rightarrow s_2$ for $r_3$ which successfully resolves the current congestion observed in Figure~\ref{fig:example}(b). 
This, however, does not increase the weights for $s_1 \rightarrow s_3$ and $s_3 \rightarrow s_2$, since these links are utilized at $30$\%, and  their weights remain at 1. Once $r_4$ arrives, OSPF directs $r_4$ to $s_1 \rightarrow s_3 \rightarrow s_2$ as it is the shortest weighted path between $s_1$ and $s_4$. This will not cause any congestion, but the weights for $s_1 \rightarrow s_3$ and $s_3 \rightarrow s_2$ increase to $w=4$ since these links are now utilized at $60$\%. Finally, OSPF will direct the last two requests $r_5$ and $r_6$ to the longer path $s_1 \rightarrow s_4 \rightarrow s_5 \rightarrow s_2$, since, now, this path is the shortest weighted path between $s_1$ and $s_2$.

As shown above, using our GP-learned link-weight formula, OSPF is now able to manage flow paths without causing any congestion and without having to re-invoke the self-adaptation planning step beyond the single invocation after the arrival of $r_3$.

We note that there are several techniques that modify network parameters at runtime to resolve congestion in SDN (e.g.,~\cite{Yashar:12,PriyadarsiniMBK19}). These techniques, however, rely on pre-defined rules. For example, to make OSPF  -- discussed above -- adaptive, a typical approach is to  define a network-weight function~\cite{RetvariNCS09}. This function can involve parameters that change at runtime and based on the state of the network. However, the structure of the function is fixed. Consequently, the function may not be suitable for all networks with different characteristics and inputs. To address this limitation, we use GP to \emph{learn and evolve the function structure} instead of relying on a fixed, manually crafted function. In Section~\ref{sec:eval}, we compare our GP-based approach with OSPF configured using an optimized weight function suggested by Cisco standards~\cite{Coltun:08,Cisco:05}. As we show there, OSPF's optimized weight function cannot address the congestion caused by the network-request bursts in our industrial case study.



In this paper, we focus on situations where congestion can be resolved by re-routing -- in other words, when the network topology is such that alternative flow paths can be created for some requests. Given our focus, OSPF is a natural comparison baseline, noting that, in OSPF, congestion is resolved by re-routing. When alternative flow paths do not exist, congestion resolution can be addressed only via traffic shaping~\cite{Yashar:12}, i.e., via modifying the network traffic. Our approach does not alter the network traffic. We therefore do not compare against congestion-resolution baselines \hbox{that use traffic shaping.}

%% file: approach.tex
\section{Self-Adaptation Control Loop}
\label{sec:selfcontrol}
\begin{figure}[t]
	\centerline{\includegraphics[width=.95\columnwidth]{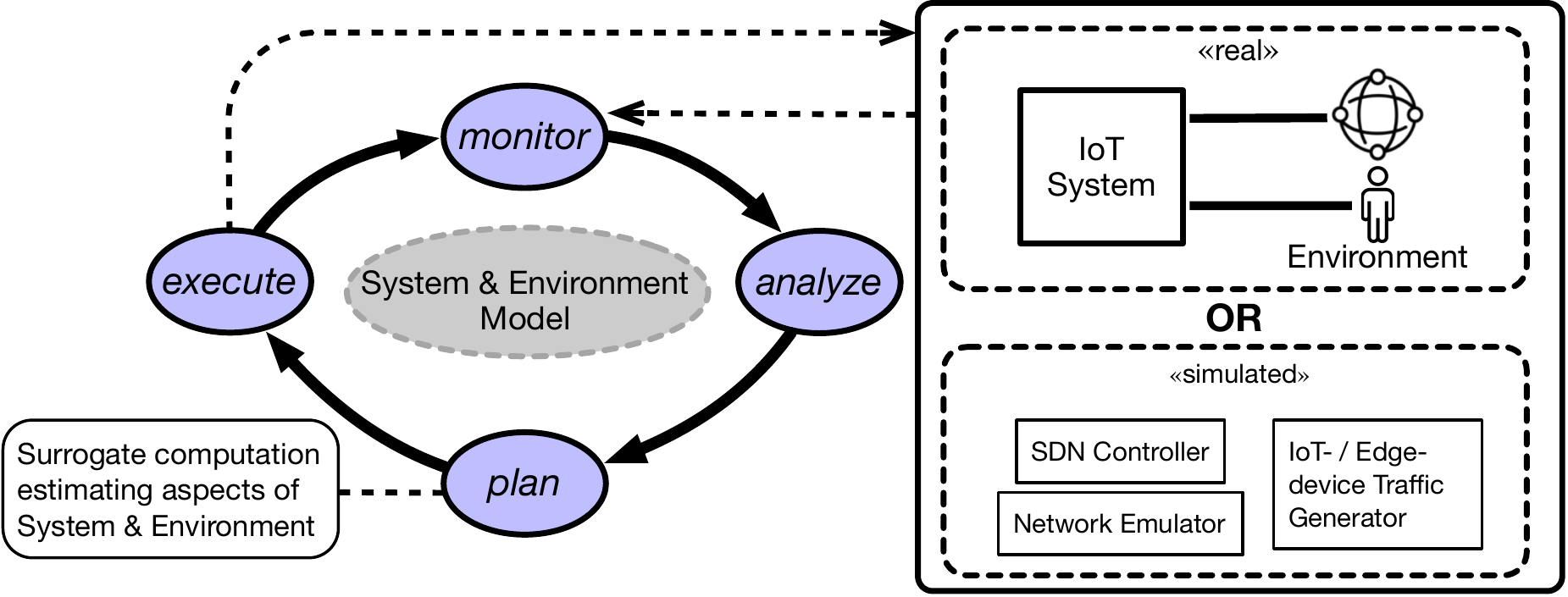}}
	\vspace*{-.2cm}
	\caption{The MAPE-K~\cite{kephart:03} self-adaptation control feedback-loop. The loop may interact with a real or simulated system. The plan step  uses surrogates to estimate dynamic aspects of the system or the environment.}\label{fig:adaptloop}
	\vspace*{-.45cm}
\end{figure}

Our approach leverages MAPE-K~\cite{kephart:03,Moreno:15}-- the well-known self-adaptation control-loop shown in Figure~\ref{fig:adaptloop}. The loop has four main steps: \emph{monitoring} the system and its environment, \emph{analyzing} the information collected from the system and its environment and deciding if adaptation is needed, \emph{planning} on how to adapt, and \emph{executing} the adaptation by applying it to the system.

Our self-adaptation loop is an add-on to the programmable control layer of the SDN architecture. In an SDN, the control layer has a global view of the entire network and can dynamically modify the network at runtime. The routing logic of an SDN is centralized, decoupled from data and network hardware, and expressed using software code at the control layer. Given the separation of control logic from data planes and forwarding hardware, changing link weights can be programmed at the control layer in a way that the changes do not affect the existing network flows and instead are used only for routing new flows~\cite{Apostolopoulos:99}. Dynamically modifying link weights therefore does not jeopardize network stability.


In our work, the self-adaptation loop interacts with a simulator instead of a real system.  Using a simulator rather than an actual system is a common approach when designing and evaluating self-adaption techniques. This is because using the actual system for this purpose is time-consuming and expensive and may further cause system wear and damage~\cite{Iftikhar:17,SotiropoulosWGI17}. In addition, simulators are highly configurable and can imitate a variety of system configurations. Evaluating self-adaptation techniques using simulators enables us to cover a multitude of systems rather than one fixed setup~\cite{paper2,paper5}. 

 As shown on the right side of Figure~\ref{fig:adaptloop}, our simulator combines three components: An SDN controller capturing the software-defined controller of a network system~\cite{Berde:14}; a network emulator that simulates the network  infrastructure including links, nodes and their properties~\cite{Lantz:10}; and a traffic generator~\cite{Botta:12} that emulates different types of requests generated by IoT devices and sensors.

We augment with GP the planning step of the self-adaptation loop. In order for GP to assess the fitness of candidate solutions, i.e., candidate link-weight functions, the system should be simulated for each candidate solution. This cannot be done using the simulator we use to replace the real system and its environment, since this simulator is real-time (wall-clock-time) and requires a few seconds to compute the behaviour of a network for each candidate weight-link formula. Our GP algorithm, when invoked, needs to explore several candidate formulas. To do so efficiently, we use \emph{surrogate} computations that approximate the fitness values~\cite{paper1,paper3,paper4}. Specifically, as we detail in Section~\ref{subsec:selfloop}, our GP algorithm computes the fitness for each candidate solution using a snapshot of the system and its environment from the simulator,  assuming that the system does not change during a small time period (expected to be less than one second,   as we explain in Section~\ref{subsec:selfloop}).

\begin{figure}[t]
\centerline{\includegraphics[width=.9\columnwidth]{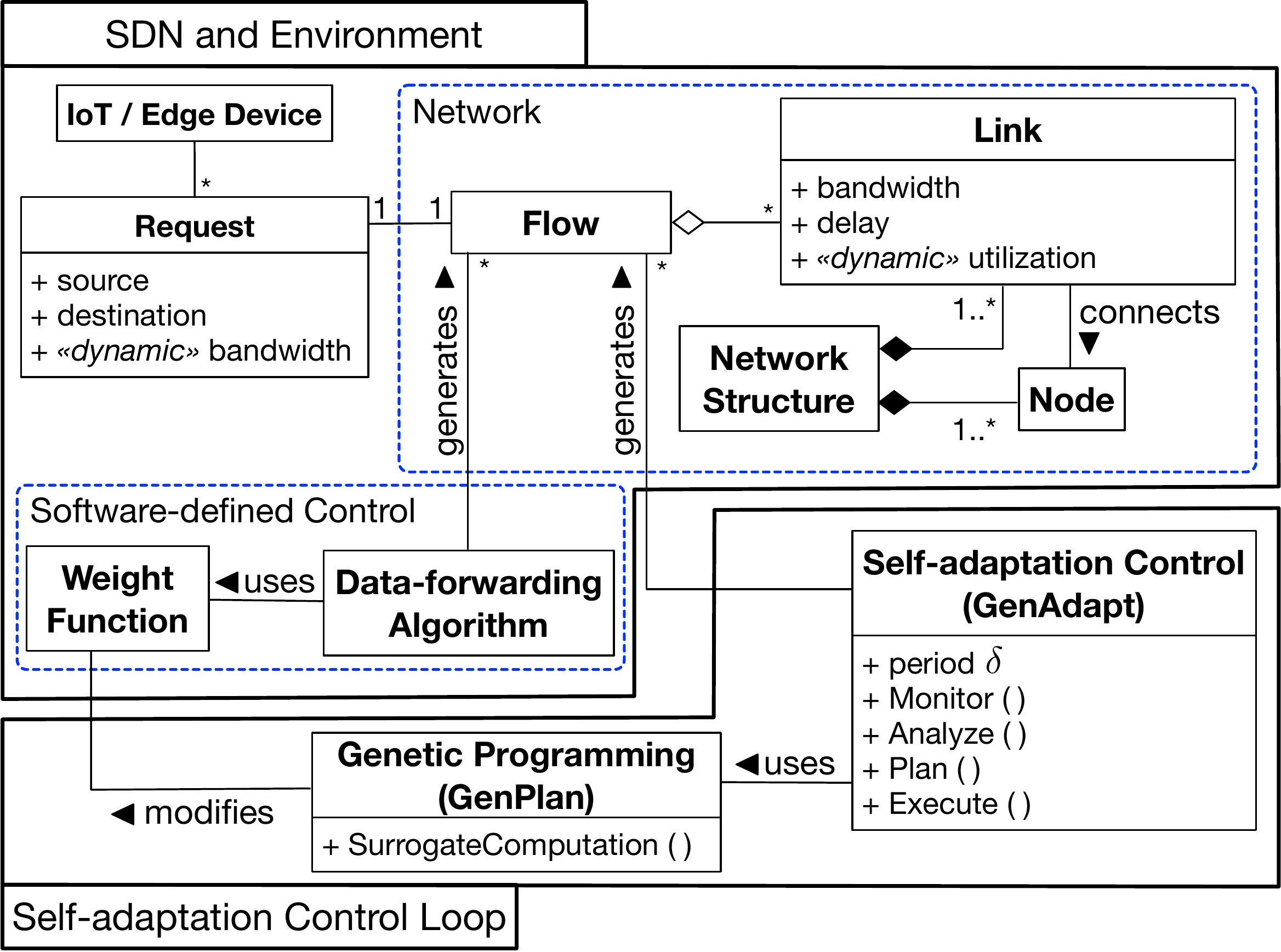}}
	\vspace*{-.2cm}
	\caption{Domain model for  self-adaptive SDN.}
	\label{fig:model}
		\vspace*{-.25cm}
\end{figure}

 Figure~\ref{fig:model} shows a domain model consisting of two packages: One package, discussed in Section~\ref{subsec:sysEnv}, captures the main elements of an SDN  and its environment, and the other, discussed in Section~\ref{subsec:selfloop}, specifies the self-adaptation control loop that we develop.

\subsection{Domain Model for  Self-adaptive SDN}
\label{subsec:sysEnv}
The \textsf{SDN and Environment} package captures the static structure of a network as well as its dynamic behaviour over time. The environment of an SDN is comprised of IoT and edge devices which are connected to the network and which generate data-transmission requests (requests, for short) that the network needs to fulfil. 

\begin{dfn}[Data-transmission Request]\label{def:datareq}
A data-transmission request $r$ specifies a data stream sent by a network node  $s$  to a network node  $t$. We denote the source node of $r$ by  $r.s$ and the destination node of $r$ by $r.t$. Let $[0..T]$ be a time interval. We denote the (data) bandwidth of $r$ at time $t \in [0..T]$ by $r.d(t)$. The bandwidth is the amount of data transmitted from $r.s$ to $r.t$ over time. We measure bandwidth in megabits per second (Mbps). The request bandwidth may vary over time (marked as ``dynamic'' in Figure~\ref{fig:model}).
\end{dfn}

In Figure~\ref{fig:model}, the network part of an SDN  is delineated with a dashed rectangle labelled \textsf{Network} and includes the \textsf{Network Structure}, \textsf{Link}, \textsf{Node} and \textsf{Flow} entities. Specifically, a network is a tuple $G=(V,E)$, where $V$ is a set of nodes, and $E \subseteq V \times V $ is a set of directed links between nodes.  Note that each (undirected) link in Figure~\ref{fig:example} represents two directed links. Network links have a nominal maximum bandwidth and  maximum transmission delay assigned to them  based on their physical features and types. The bandwidth of a link is the maximum capacity of the link for transmitting data per second, and the maximum delay specifies the maximum time it takes for data transmission over a link.


\begin{dfn}[Static Properties of a Link]\label{def-statLn}
Let $G = (V, E)$ be a network structure. Each link $e \in E$ has a bandwidth $bw(e)$ and a nominal delay $dl(e)$. We denote by $\mathit{StaticProp}(e)$ the tuple $(bw(e), dl(e))$, indicating the static properties of $e$.  
\end{dfn}

As discussed in Section~\ref{sec:motivate}, a network  handles requests by identifying  a directed path (or a \emph{flow}) in the network to transmit them. Upon the arrival of each request $r$, a network flow (path) $f$ is established to transmit the data stream of $r$ from the requested source $r.s$ to the requested destination $r.t$. Each flow $f$ is a directed path of links that connects $r.s$ to  $r.t$. As shown in Figure~\ref{fig:model},  one flow is created per request. The bandwidth of each flow is equal to that of its corresponding request. Since requests have dynamic bandwidths  (Definition~\ref{def:datareq}), flows have dynamic bandwidths too. We define the throughput of link $e$ at time $t$, denoted by $\mathit{throughput}(e, t)$, as the total of the bandwidths of the flows going through $e$ at time $t$. 

\begin{dfn}[Dynamic Utilization of a Link]\label{def-linkdyn}
Let $G = (V, E)$ be a network, and let $[0..T]$ be a time interval. At each time $t \in [0..T]$, each network link $e \in E$ has a utilization $\mathit{util}(e, t)$ which is computed as follows:
$\mathit{util}(e, t) = \mathit{throughput}(e, t)/\mathit{bw}(e)$.
\end{dfn}




The software-defined control entities  in Figure~\ref{fig:model} include \textsf{Weight Function} and \textsf{Data-forwarding Algorithm}. Data-forwarding algorithms are event-driven and handle requests upon arrival. As discussed in Section~\ref{sec:motivate}, data forwarding typically generates flows based on shortest weighted paths between the source and the destination of a request. Thanks to the SDN architecture, link weights are programmable and can be computed by a weight function that accounts for both the dynamic and the static properties of networks. For example, the GP-derived link-weight function in Figure~\ref{fig:example} (c)  uses the dynamic link utilization  $\mathit{util}(e, t)$ and the static parameter $\mathit{threshold}$ to compute link weights.

%% file: algorithmGP.tex
\subsection{Generative Self-adaptation}
\label{subsec:selfloop}
In this section, we describe \textit{GenAdapt}, our   self-adaptation control loop, and \textit{GenPlan}, the genetic algorithm used in the planning step of GenAdapt. As discussed earlier,  self-adaptation control has four main steps; these are specified as methods in the self-adaptation control entity in Figure~\ref{fig:model}. GenAdapt, i.e., our self-adaptation loop, runs in parallel with the data-forwarding algorithm. In contrast to the  data-forwarding algorithm which is event-driven, GenAdapt is executed periodically with a period $\delta$,  indicated as an attribute of GenAdapt in Figure~\ref{fig:model}.  The self-adaptation loop periodically monitors the network for congestion as the environment changes, e.g., due to the arrival of new requests.  There is a trade-off between the execution time of GenAdapt and the period $\delta$. In particular, $\delta$ should be small enough so that GenAdapt is executed frequently to detect and handle congestion promptly. At the same time, $\delta$ should be large enough so that frequent executions of GenAdapt do not interfere with other SDN algorithms and applications~\cite{ShinNSB0Z20}. 

\begin{figure}[t]
\begin{lstlisting}[style=Alg]
Algorithm GenAdapt
Input G : Network structure
Input $\cup_{e \in E}$StaticProp($e$): Static properties of links
Input BestSol: Best Solutions from the previous round
Output F: Optimized flows
Output W: Optimized weight function

for every time step $i$ $\in \{1, \ldots, n\}$ do 
?\vrule? F$_i$, $\cup_{e \in E}$util($e$, $i\cdot \delta$) $\leftarrow$ Monitor() //Dynamic data from SDN Sim
?\vrule? maxUtil = $\mathit{Max}\{\mbox{util}(e, i\cdot \delta)\}_{e \in E}$ 
?\vrule? if maxUtil $>$  $\mathit{threshold}$  then
?\vrule? ?\vrule? F, W $\leftarrow$ GenPlan(G,  $\cup_{e \in E}$StaticProp($e$), F$_i$, BestSol)
?\vrule? ?\vrule? Apply F $\mbox{and}$ W to the SDN data-forwarding algorithm
?\vrule? $\mathbf{end}$
$\mathbf{end}$
\end{lstlisting}
\vspace*{-.25cm}
\caption{Self-adaptation loop to resolve congestion by learning a new weight function for SDN data-forwarding.}
\label{alg:self-adapt}
\vspace*{-.45cm}
\end{figure}

GenAdapt, shown in Figure~\ref{alg:self-adapt}, executes at every time step $i{\cdot} \delta$ ($i=0, 1, \ldots$).  The \emph{monitor} step (line 10) fetches the set $F_i$ of flows at time $i \cdot \delta$ and the utilization $\mathit{util}(e, i\cdot \delta)$ of every  link $e$ at $i \cdot \delta$. The \emph{analyze} step (lines 11-12) determines  whether the network is congested, i.e., whether adaptation is needed. A network is congested if there is a link $e \in E$ which is utilized above a certain threshold~\cite{Lin:16,Akyildiz:14}. To detect congestion, the maximum utilization of all the links is compared with $\mathit{threshold}$, which is a fixed parameter of the SDN. 

If the network is congested, GenAdapt calls \textit{GenPlan} (line 13). GenPlan, shown in Figure~\ref{fig:gp}, is our GP algorithm which we discuss momentarily. The output of GenPlan is a new  link-weight function as well as a modified set of flows where a minimal number of flows have been re-routed.  The new flows and the optimized link-weight function are then applied to the SDN under analysis (line 14).  Note that the only  change that needs to be done to the network is re-routing a typically small number of flows to eliminate congestion. 


 GenPlan (Figure~\ref{fig:gp}) generates link-weight functions that optimize a fitness function characterizing the desired network-flow properties. The link-weight functions are specified  in terms of static link properties ($\mathit{bw}(e)$ and $\mathit{dl}(e)$), dynamic link utilization ($\mathit{util}(e,t)$) and utilization threshold. Note that GenPlan always reuses in its initial population half of the best solutions (candidate weight functions) generated by its previous invocation and stored in a variable named BestSol. \emph{Reusing as bootstrap some of the best solutions from the previous round ensures continuity and incrementality in learning the weight functions across multiple rounds of self-adaptation}.

 GenPlan  starts by selecting a number of flows, BadFlows, from a congested set of flows, OldFlows, using the FindFlowsCausingCongestion algorithm shown in Figure~\ref{fig:removeflows}. BadFlows are the flows that should be re-routed to resolve congestion. Following the standard steps of GP,  GenPlan creates an initial population $P_0$ (line 14) containing a set of possible weight functions (individuals). For every individual $\omega \in P_0$, we call ComputeSurrogate, shown in Figure~\ref{fig:surrogate}, to re-route the flows in BadFlows when $\omega$ is used to compute weights of the network links (line 19 of GenPlan). Specifically, for each  $\omega$,   ComputeSurrogate  generates the set NewFlows which includes (1)~the flows corresponding to those in BadFlows, but re-routed based on $\omega$; and (2)~the original flows that did not cause congestion (i.e., OldFlows$\setminus$BadFlows). The set NewFlows is needed to compute the fitness value for $\omega$ (line 21).  The fitness function aims to determine how close an individual is to resolving congestion. Our fitness function combines three criteria as we elaborate momentarily. GenPlan evolves the population by breeding and generating a new offspring population (line 17). The breeding and evaluation steps are repeated until a stop condition is satisfied. Then, GenPlan  returns the link-weight function with the lowest fitness (BestW) and its corresponding flows (line 28).

The ComputeSurrogate algorithm (Figure~\ref{fig:surrogate}), which is called by GenPlan, estimates the flows for each candidate weight function $\omega$, assuming that the flow bandwidths obtained at the monitoring step are constant over the duration of $\delta$. ComputeSurrogate first updates the utilization and weight values for each link, assuming that BadFlows are absent (lines 9-12). The new weights are then used to re-route each flow $f$ in BadFlows by identifying a flow $f'$ as the new shortest path (line 14). After creating  each new flow $f'$, the utilization and weight values for each link on  $f'$ are updated (lines 15-18). Note that, because of the assumption that flow bandwidths remain fixed during $\delta$, the utilization-per-link $util(e)$ in ComputeSurrogate is not indexed by time.  


\begin{figure}[t]
\begin{lstlisting}[style=Alg]
Algorithm GenPlan
Input G: Network structure
Input $\cup_{e \in E}$StaticProp($e$): Static properties of links
Input OldFlows: Current network flows
Input BestSol: Best Solutions from the previous round
Output W: New weight function 
Output NewFlows: New network flows 

t = 0;
BadFlows = FindFlowsCausingCongestion(OldFlows, G);
Flows = OldFlows $\setminus$ BadFlows
while not(stop_condition) do 
?\vrule? if (t == 0)
?\vrule? ?\vrule? P$_0$ = InitialPopOfWeightFormulas() $\cup$ BestSol
?\vrule? ?\vrule? OffSprings = P$_0$
?\vrule? else
?\vrule? ?\vrule? OffSprings = Breed(P$_t$)
?\vrule? $\mathbf{end}$
?\vrule? for $\omega \in$ OffSprings do
?\vrule? ?\vrule? NewFlows = ComputeSurrogate(Flows, BadFlows, G, $\omega$) 
?\vrule? ?\vrule? $\omega$.Fit = Evaluate (G, NewFlows, OldFlows)
?\vrule? $\mathbf{end}$
?\vrule? P$_{t+1}$ = OffSprings
?\vrule? t = t + 1
$\mathbf{end}$
BestW = BestSolution(P$_0$, $\ldots$, P$_t$)
bestFlows = ComputeSurrogate(Flows, BadFlows, G, BestW) 
return BestW, bestFlows
\end{lstlisting}
\vspace*{-.25cm}
\caption{Generating a new  link-weight function and a congestion-free set of flows using Genetic Programming.}
\label{fig:gp}
\vspace*{-.42cm}
\end{figure}

Following standard practice for expressing meta-heuristic search problems~\cite{Harman2010SearchBS}, we define the representation, the fitness function, and the genetic operators underlying GenPlan.

\textbf{Representation of the Individuals.} An individual represents some weight function induced by the following grammar rule:\\[0em]

\hspace*{-.8em}\fbox{\hspace*{-.5em}\scalebox{.85}{\tt\begin{tabular}{l@{\ }l@{\ } l@{\ } p{0.05mm} l@{\ } l@{\ } p{63mm}}
        exp &::= &
    exp\,($+\mid-\mid*\mid/$)\,exp\,$|$\,const\,$|$\,StaticProp\,$|$\,DynamicVar\,$|$\,param\\
\end{tabular}}\hspace*{-.5em}}\\[0em]

In the above, the symbol $\mid$ separates alternatives, \texttt{const} is an ephemeral random constant generator~\cite{Veenhuis13},  \texttt{StaticProp} are static link properties (Definition~\ref{def-statLn}), \texttt{DynamicVar} is the link utilization  (Definition~\ref{def-linkdyn}), and \texttt{param} is some network parameter.  
The formula in Figure~\ref{fig:example}(c) can be generated by this grammar rule and is thus an example individual in GenPlan. Note that the SDN data-forwarding algorithm assumes that link values are integers. The $\lfloor \rfloor$ used in the formula of Figure~\ref{fig:example}(c) rounds down the output to an integer. Otherwise, $\lfloor \rfloor$ is not part of the weight formulas.

\begin{wrapfigure}[6]{r}{3cm}
\vspace*{-.6cm}
 \hspace*{-.4cm} \centerline{\includegraphics[width=0.4\columnwidth]{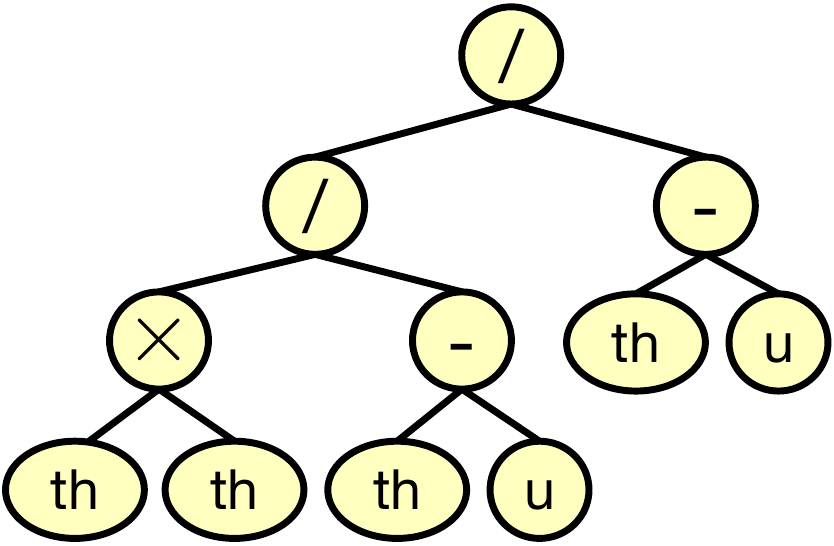}}
\end{wrapfigure} 
Each individual is constructed and 
manipulated as a parse tree. For example, the figure to the right shows the parse tree corresponding to $\frac{th^2}{ (th - u)^2}$ where \textsf{u} is a shorthand for $util(e, t)$ and \textsf{th} is a shorthand for threshold. The initial population of GenPlan is generated by randomly building parse trees using the grow method~\cite{poli2008field} (i.e., the root and inner nodes are labelled by the mathematical operations, and the leaves are labelled by variables, constants or parameters specified in the grammar). As discussed earlier when GenPlan is called for the first time,  the initial population is generated randomly. For subsequent calls to GenPlan, half of the initial population is generated and the other half is reused from the best elements in the last population generated by the previous invocation of GenPlan.

\begin{figure}[t]
\begin{lstlisting}[style=Alg]
Algorithm FindFlowsCausingCongestion
Input G: Network structure
Input Flows: Current flows
Output BadFlows: A subset of Flows causing congestion 

BadFlows = $\emptyset$
while G is congested do 
?\vrule? Let $e$ $\mbox{be}$ the most congested link
?\vrule? Let $f \in$ Flows such that  $e \in f$ // selected randomly
?\vrule? BadFlows = BadFlows$\cup \{f\}$
?\vrule? Flows = Flows$\setminus \{f\}$
$\mathbf{end}$
return BadFlows
\end{lstlisting}
\vspace*{-.25cm}
\caption{Selecting a subset of flows whose removal will resolve congestion.}
\label{fig:removeflows}
\vspace*{-.1cm}
\end{figure}

\begin{figure}[t]
\begin{lstlisting}[style=Alg]
Algorithm ComputeSurrogate 
Input G: Network structure
Input Flows: Current congestion-free set of flows
Input BadFlows: Flows that have to $\mbox{be}$ re-routed 
Input $\omega$: Candidate weight function
Output NewFlows: BadFlows re-routed using $\omega$

NewFlows $= \emptyset$
for $e \in E$ do
?\vrule? util($e$) $\leftarrow$ sum of the bandwidths of $f \in$ Flows s.t. $e \in f$
?\vrule? Compute weight $\mbox{for}$ $e$ using $\omega$
$\mathbf{end}$
for $f \in$ BadFlows do
?\vrule? $f'$ $\leftarrow$ shortest weighted path from  source to destination of $f$ 
?\vrule? for $e \in f'$ do
?\vrule? ?\vrule? util($e$) = util($e$) + bandwidth of $f'$ 
?\vrule? ?\vrule? update the weight $\mbox{of}$ $e$ using $\omega$
?\vrule? $\mathbf{end}$
?\vrule? NewFlows = NewFlows$\cup \{f'\}$
$\mathbf{end}$
return NewFlows $\cup$ Flows
\end{lstlisting}
\vspace*{-.25cm}
\caption{Re-routing congested flows (BadFlows) by computing shortest paths based on a candidate weight function ($\omega$).}
\label{fig:surrogate}
\vspace*{-.4cm}
\end{figure}

\textbf{Fitness Function.}
Our fitness function is hybrid and  combines the following three metrics: (1)~Maximum link utilization across all the network links ($\mathit{Fit1}$).
For each individual weight function $\omega$, GenPlan computes the set NewFlows of flows based on link weights generated by $\omega$.  The $\mathit{Fit1}$ metric computes the utilization value of the most utilized network link, considering the flows in NewFlows. If $\mathit{Fit1}$ is higher than the threshold, then the network is congested. Hence, we are interested in individuals whose $\mathit{Fit1}$ is less than the threshold. (2)~The cost of re-routing network flows measured as the number of  link updates, i.e., insertions and deletions, required to reconfigure the network flows ($\mathit{Fit2}$). In GenPlan, OldFlows is the current set of congested flows, and as mentioned above, NewFlows is the set of flows computed for each individual $\omega$. We compute $\mathit{Fit2}$ as the edit distance between each pair of flows $f \in$ OldFlows  and $f'\in$ NewFlows  such that $f$ and $f'$ are both related to the same request.  Specifically, the distance between two flows $f$ and $f'$ is measured as the longest common  subsequence (LCS) distance of two paths~\cite{Cormen:09} by counting the number of link insertions and link deletions required to transform $f$ into $f'$. We note that this metric has previously been used as a proxy for the reconfiguration cost of network flows~\cite{ShinNSB0Z20}. (3)~The total data transmission delay generated by the new flows ($\mathit{Fit3}$). The $\mathit{Fit3}$ metric is computed as the sum of all the delay values, i.e., $\mathit{dl(e)}$, of the links that are utilized by the new flows (i.e., the NewFlow set).  The larger this value, the higher the overall transmission delay induced by NewFlows. The last metric allows us to penalize candidates that generate longer flows compared to those that generate shorter ones. 

The $\mathit{Fit1}$, $\mathit{Fit2}$ and $\mathit{Fit3}$ metrics have different units of measure and ranges. Thus, before combining them, we normalize them using the well-known rational function $\bar{x} = x/(x + 1)$. This function provides good guidance to the search for minimization problems compared to other alternatives.

Among the three metrics, lowering  $\mathit{Fit1}$ below the congestion threshold takes priority; if a candidate solution is unable to resolve congestion, then we are not interested in the other two metrics. Once $\mathit{Fit1}$ is below the threshold, we do not want to lower $\mathit{Fit1}$ any further, since we want the network optimally utilized but not congested. Instead, we are interested in lowering the cost and delay metrics.  We denote the normalized forms of our three metrics by $\overline{\mathit{Fit1}}$, $\overline{\mathit{Fit2}}$ and $\overline{\mathit{Fit3}}$, respectively, and define the following overall fitness function to combine the three: 
\vspace*{.4em}

$\begin{array}{l}
Fit =
\begin{cases}
\overline{\mathit{Fit1}} +  2 &  \text{(1) If $\mathit{Fit1} \geq \mathit{threshold}$}\\
\overline{\mathit{Fit2}} + \mathit{\overline{Fit3}} &  \text{(2) If  $\mathit{Fit1} < \mathit{threshold}$} \\
\end{cases} 
\end{array}$
\vspace*{.4em}

Given a candidate weight function $\omega$,  evaluating $\mathit{Fit}(\omega)$ always yields a value in $[0..3]$:  $\mathit{Fit}(\omega) \geq 2$ indicates that $\omega$ is not able to resolve congestion; and $\mathit{Fit}(\omega) <2 $ indicates that $\omega$ can resolve congestion, and its fitness determines how well  $\omega$ is doing in reducing  cost and delay. Note that in our fitness function defined above, cost and delay are equally important; we do not prioritize either one. If desired, one can modify the above function by adding coefficients to $\overline{\mathit{Fit2}}$  and $\overline{\mathit{Fit3}}$ to prioritize cost over delay or vice versa.

\textbf{Genetic Operators.}  
We use one-point crossover~\cite{onepointcrossover}. It randomly  selects  two  parent individuals. It then randomly selects one sub-tree in each parent, and  swaps the  selected  sub-trees  resulting  in  two  children. For the mutation operator, we  use  one-point  mutation~\cite{poli1998schema}  that mutates a child individual by randomly selecting one sub-tree and replacing it with a randomly generated tree, which is generated using the initialization procedure. For the parent selection operator, we use tournament selection~\cite{Luke:13}.

%% file: eval.tex
\section{Empirical Evaluation}
\label{sec:eval}
In this section, we investigate the following research questions (RQs) using open-source synthetic and industrial IoT networks. To answer RQ1, we use the industrial network and ten synthetic networks. To answer RQ2, we use eight more synthetic networks alongside two of the synthetic networks from RQ1. In total, our evaluation involves one industrial network and  18 synthetic ones.  

\textbf{RQ1 (Effectiveness)} \emph{How effective is GenAdapt in modifying the logic of the SDN data-forwarding algorithm to avoid future congestions?} 
The main novelty of GenAdapt is in attempting to adapt the logic of SDN data forwarding (i.e., the link-weight functions) instead of adapting its output (i.e., the individual flow paths). Through RQ1, we compare GenAdapt with two baseline techniques: (i)~an approach, named DICES~\cite{ShinNSB0Z20}, which, similar to GenAdapt, uses MAPE-K self-adaptation, but optimizes individual flow paths; and (ii)~OSPF  configured using a standard heuristic for setting optimized link weights~\cite{Coltun:08,Cisco:05}. By comparing GenAdapt with these baselines, we investigate \emph{whether GenAdapt, when called to resolve an existing congestion, is able to reduce the number of occurrences of congestion in the future, without incurring additional overhead.}

\textbf{RQ2 (Scalability):} \emph{Can GenAdapt resolve congestion efficiently as the size of the network and the number of requests increase?} To assess scalability, we evaluate the execution time of GenAdapt as the size of the network and the number of  requests increase. 

\textbf{Implementation and Availability.}
We implemented GenAdapt using ECJ (version 27)~\cite{DBLP:conf/gecco/ScottL19} and the open-source version of DICES~\cite{ShinNSB0Z20}, which implements a self-adaptive control loop over the Open Network Operating System (ONOS)~\cite{Berde:14} -- the SDN controller that we employ in our work. Similar to DICES, our simulator consists of three open-source tools: (i)\,ONOS~\cite{Berde:14} as the ``SDN Controller'' (see Figure~\ref{fig:adaptloop}), (ii)\,Mininet~\cite{Lantz:10} as the ``Network Emulator'',  and (iii)\,Distributed Internet Traffic Generator (D-ITG)~\cite{Botta:12} as the ``IoT-/Edge-device Traffic Generator''. Our simulator is memory-intensive and  runs on two Linux virtual machines (VMs), but GenAdapt by itself does not require special considerations for memory or CPU power. All experiments were performed on a machine with a 2.5 GHz Intel Core i7 CPU and 32 GB of memory. Our implementation and all experimental material are available online~\cite{appedix}.


\subsection{RQ1 -- Effectiveness}
\label{sec:rq1}
Before answering RQ1, we present the baselines, the study subjects,  the configuration of GenAadapt and the setup of our experiments. 



\textbf{Baselines.}  Our first baseline,  DICES, is from the SEAMS literature. Similar to GenAadapt, DICES is self-adaptive, but unlike GenAadapt, it uses a genetic algorithm to modify the individual flows generated by the SDN data-forwarding algorithm.  The second baseline is  OSPF~\cite{Coltun:08,Cisco:05} configured by setting the link weights to be inversely proportional to the bandwidths of the links as suggested by Cisco~\cite{Cisco:05}. The OSPF heuristic link weights are meant to induce optimal flows that eliminate or minimize the likelihood of congestion. OSPF is widely used in real-world systems~\cite{FortzT02} and as a baseline in the literature~\cite{Poularakis:19,Amin:18,Bianco:17,Rego:17,Caria:15,Bianco:15,Agarwal:13}.



\textbf{Study Subjects.} For RQ1, we use: (1)~ten synthetic networks, and (2)~an industrial SDN-based IoT network published in earlier work~\cite{ShinNSB0Z20}. For the synthetic networks, we consider two  network topologies: (i)~complete graphs, and (ii)~multiple non-overlapping paths between node pairs. The network in Figure~\ref{fig:example} is an example of the latter topology, where $s_1$ and $s_2$ are connected by three non-overlapping paths.
Since our approach works by changing flow paths, we naturally focus our evaluation on topologies with multiple paths from a source to a destination, thus excluding topologies where adaptation through re-routing is not possible. The topologies that we experiment with, i.e., (i) and (ii) above, are the two extreme ends of the spectrum in terms of path overlaps between node pairs: In the first case, we have complete graphs where there are as many  overlapping paths as can be between node pairs; and, in the second case, we have no overlapping paths at all. We consider four complete graphs -- referred to as FULL hereafter -- with five, six, seven and ten nodes, and consider two multiple-non-overlapping path graphs -- referred to as MNP hereafter -- with five and eight nodes. More precisely, one MNP graph has five nodes connecting the designated source and destination with three non-overlapping paths (Figure~\ref{fig:example}); the other has eight nodes doing the same with four (non-overlapping) paths. Following the suggested parameter values in the existing literature for such experiments~\cite{ShinNSB0Z20}, we set the static properties of the links, namely bandwidth and delay, to 100Mbps and 25ms, respectively.

To instigate changes in the SDN environment, we generate data requests over time and not at once. We space the requests $10$s apart to ensure that all the requests are properly generated and that the network has some time to stabilize, i.e.,  we send requests at $0$s, $10$s, $20$s, $30$s, and so on. For each FULL network, we fix a source and a destination node, and  generate either three or four requests every $10$s. For each MNP network, we generate every $10$s two requests between the end-nodes connected by multiple paths. The reason why we generate fewer requests in the MNP networks is because there are fewer paths compared to the FULL networks. For a given network (FULL or MNP), the generated requests have the same bandwidth. The request bandwidth for each network is selected such that  some congestion is created starting from $10$s. The request bandwidths are provided online~\cite{appedix}. We denote our synthetic networks by \hbox{FULL($x$, $y$)} and MNP($x$, $y$), where $x$ is the number of nodes and $y$ is the number of requests generated every $10$s. The ten synthetic network that we use in RQ1 are four FULL networks, once with three requests and once with four requests generated every $10$s, and two MNP networks with two requests generated every $10$s.

Our industrial subject is an emergency management system (EMS)  from previous work~\cite{ShinNSB0Z20}. EMS represents a real-world application of SDN in a complex IoT system. This subject contains seven switches and 30 links with different values for the links' static properties. The EMS subject includes a traffic profile characterizing anticipated traffic at the time of a natural disaster (e.g., flood), leading to congestion in the network of the monitoring system. In particular, the network is used for transmitting 28 requests capturing different data stream types such as audio, video and sensor and map data. The static properties of the network links and the data-request sizes are available online~\cite{appedix} \footnote{Due to platform differences, we could not reproduce the congestion reported in~\cite{ShinNSB0Z20}. So, we increased the request bandwidths by 30\% to reproduce congestion.}. 

\textbf{Experiments and metrics.} RQ1 has two goals: (G1)~determine whether, by properly modifying the link-weight function, GenAdapt is able to reduce the number of times congestion happens, and (G2)~determine whether GenAdapt can respond quickly enough so that prolonged periods of congestion can be avoided.

The synthetic subjects are best used for achieving G1, since the requests in these subjects are generated with time gaps as opposed to all at once, which is the case in EMS (industry subject). This characteristic of the synthetic subjects creates the potential for congestion to occur multiple times, in turn allowing us to achieve G1. For G1, we compare GenAdapt with DICES; comparison with the OSPF baseline does not apply, since OSPF's heuristic cannot avoid congestion for our synthetic subjects, and neither can it resolve congestion. To compare GenAdapt with DICES, we keep track of how many times congestion occurs during our simulation, the execution time of each technique to resolve each congestion occurrence, and the total time during which the network is in a congested state. In addition, we measure packet loss, which is a standard metric for detecting congestion in network systems. Packet loss is measured as the number of dropped packets divided by the total number of packets in transit across the entire network during simulation.  The simulation time for each synthetic subject was set to end $10$s after the generation of the last request. Recall that, in our synthetic subjects,  the requests are sent at intervals of $10$s and as long as the number of paths in the underlying network is sufficient to fulfill the incoming the requests.

To achieve G2, we use EMS in order to compare GenAdapt with DICES and OSPF in terms of handling the congestion caused by requests arriving at once. Since DICES and GenAdapt are self-adaptive, they monitor EMS and attempt to resolve congestion when it occurs.  In the case of OSPF, however, the heuristic link weights are meant to reduce the likelihood of congestion and packet loss.  For this comparison, we report the total packet loss recorded by each of the three technique over a fixed simulation interval of $5$min, and also whether, or not,  GenAdapt and DICES were able to resolve the congestion in EMS within the simulation time.



\textbf{Configuring GenAdapt.} Table~\ref{tab:parameters} shows the configuration  parameters used for GenAdapt. For the mutation and crossover rates, the maximum tree depth and the tournament size, we chose  recommendations from either the GP literature~\cite{poli2008field,LukeP06} or ECJ's documentation~\cite{DBLP:conf/gecco/ScottL19}. We set $\delta$ to $1$s since this is the smallest monitoring period permitted by our simulator. We use the utilization threshold given in the literature~\cite{ShinNSB0Z20,Akyildiz:14,Lin:16}.  Since the range for link-utilization values is [$0$\% .. $100$\%], we set the minimum and maximum of the constants in GP individuals to $0$ and $100$, respectively. 

\begin{table}
\caption{Parameters of GenAdapt.}
\label{tab:parameters}
\vspace*{-1.5em}
\begin{center}
\scalebox{.7}{
\begin{tabular}{p{3.5cm} p{0.4cm}||p{6cm} p{0.4cm}}
\toprule
Mutation rate  &  0.1 & Utilization threshold  &  $80$\% \\
\hline
Crossover rate &  0.7 & Minimum of the constant in the GP grammar &   0 \\
\hline
Maximum depth of GP tree &  5 & Maximum of the constant in the GP grammar   &  100 \\
\hline
Tournament size &  7  &  Interval between invocations of GenAdapt ($\delta$) &  $1$s   \\
\hline

Population size &  10 & 
\\
\bottomrule
\end{tabular}}
\vspace*{-.3cm}
\end{center}
\end{table}

To determine the population size and the number of generations for GP, we note that, ideally, GenAdapt should not take longer than $\delta$ to execute. The execution time of GenAdapt is likely shorter when it is called in the first rounds of request generation than in the later rounds, since there are fewer flows in the early rounds of our synthetic subjects. Hence, using the fixed time limit of $1$s to stop GenAdapt is not optimal. Instead of using a time limit, we performed some preliminary experiments on our synthetic subjects to configure population size and the number of generations for the two topologies of FULL and MNP. For both topologies,  we opt to use a small population size (i.e., $10$). As for the number of generations, for the FULL networks, we stop GenAdapt when the fitness function falls below two (i.e., when congestion is resolved but the solution is not necessarily optimized for delay and cost) or when 200 generations is reached. This will ensure that GenAdapt's execution time does not exceed $1$s for our FULL networks. The MNP networks are, however, sparser and we are able to increase the number of generations while keeping the execution time below $1$s. Specifically, for MNP networks with five nodes, we use 300 generations; and, for the one with eight nodes we use 500 generations.  For EMS, the topology is more similar to a full graph, and as such, we use the same configuration for EMS as that for FULL networks. 

%% file: results.tex
\begin{figure}[t]
	\centerline{\includegraphics[width=.95\columnwidth]{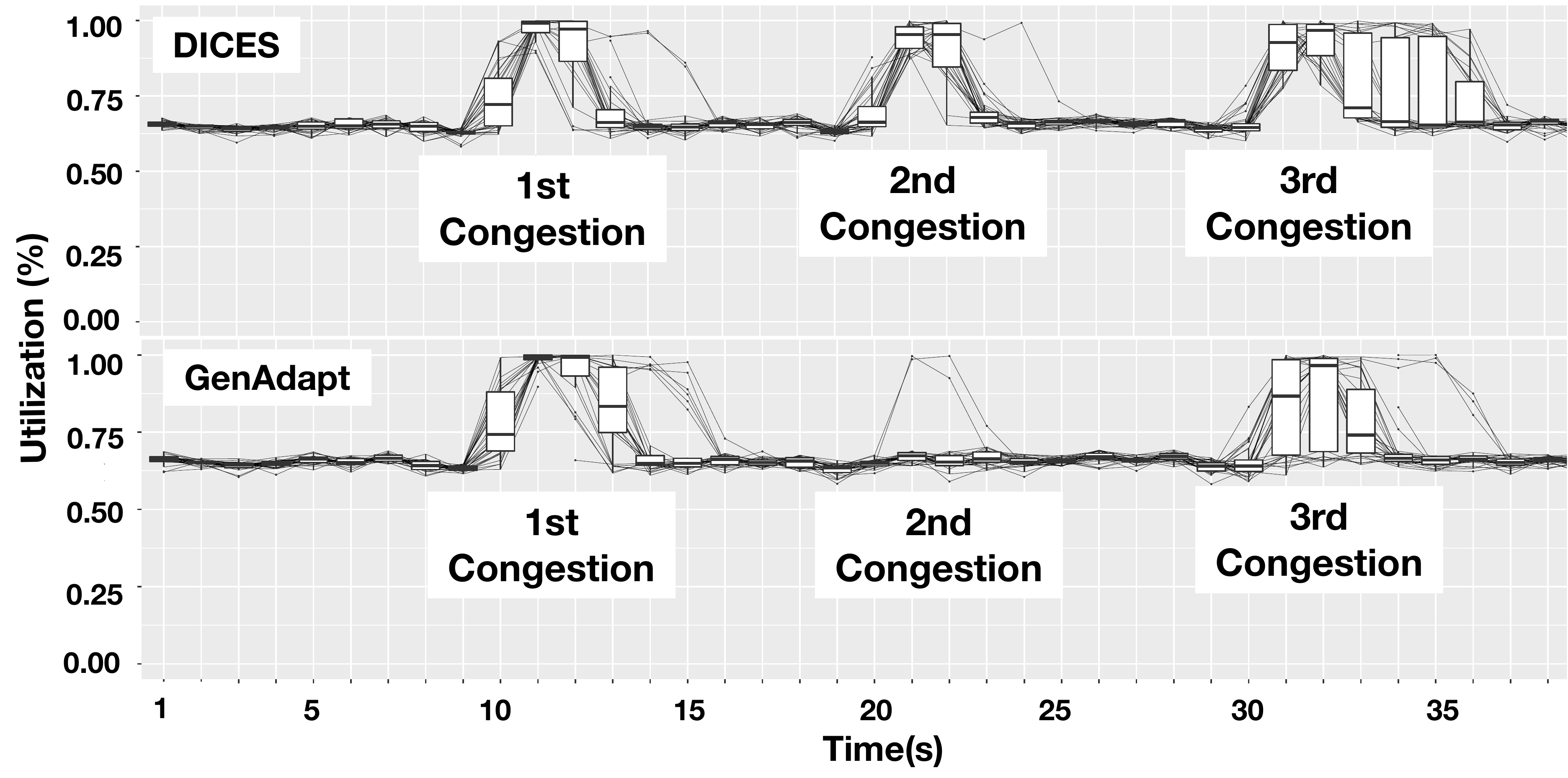}}
	\vspace{-1.2em}
	\caption{Comparing network utilization values over time obtained from 30 runs of DICES and GenAdapt for resolving congestion in an example synthetic subject: MNP(8,2).}
	\label{fig:rq1}
		\vspace{-.4cm}
\end{figure}

\textbf{Results.} 
As an example, Figure~\ref{fig:rq1} shows the network utilization over time when GenAdapt and DICES are used to resolve congestion for the synthetic network MNP(8, 2). Two simultaneous network requests are sent at 0s, 10s, 20s and 30s, leading to congestion at 10s, 20s, and 30s.  The network is congested when the utilization value is above 0.8 (i.e., the utilization threshold). Out of 30 runs for DICES, 18 runs record three congestion occurrences and 12 runs record four. In contrast, out of 30 runs for GenAdapt, six runs record only one congestion occurrence, 16 runs record two, seven runs record three, and only one run records four congestion occurrences.  Note that three congestion occurrences at 10s, 20s, and 30s are visible in the figure. However, in the simulated environment, similar to the physical world, there are some small time gaps between the arrivals of the two requests generated each time, and in addition, there are small fluctuations in flow bandwidths over time. Hence, the monitoring step of GenAdapt or DICES may detect congestion twice (i.e., once per request arrival), instead of only once and after the arrival of both requests. In total, for the example in Figure~\ref{fig:rq1}, the average number of congestion occurrences is 2.1 with GenAdapt and 3.4 with DICES. In addition, with DICES, the average of the total time that the network remains congested  is  7.6s, while with GenAdapt, this is reduced to 5.57s.

\input{table1}


With the analysis for RQ1 intuitively explained over one subject, namely MNP(8,2), we now present the complete results for this RQ. Table~\ref{tab:GpvsDices} compares GenAdapt against DICES for our ten synthetic subjects (i.e., four FULL($n$, 3), four FULL ($n$, 4) and two MNP($n$, 2) where $n$ is the number of network nodes) by reporting the average number  of congestion occurrences, the average total duration that the network is congested, the average execution time, and the average packet-loss values obtained by 30 runs of DICES and GenAdapt. For four subjects, some runs of DICES failed to resolve the last congestion within the simulation time. Specifically, for FULL(6,3), 12 runs; for FULL(6,4), 10 runs; for FULL(7,3), 20 runs; and for FULL(10,4), 3 runs of DICES failed to resolve congestion. For these subjects, the reported average number of congestion occurrences and execution times capture only the successful runs of DICES. In contrast, all runs of GenAdapt for all the subjects successfully resolved congestion within the simulation time.

We compare the results of Table~\ref{tab:GpvsDices} through statistical testing. We use the non-parametric pairwise Wilcoxon rank sum test~\cite{capon:91} and the Vargha-Delaney's $\hat{A}_{12}$ effect size~\cite{vargha:00}.
We first focus on the following three metrics: number of congestion occurrences, congestion duration, and packet loss. For fives subjects -- MNP(8,2), MNP(5,2), FULL(10,4), FULL(6,4), and FULL(6,3) -- the $p$-values for all the comparisons of the above three metrics are lower than $0.05$ and the $\hat{A}_{12}$ statistics show large or medium effect sizes, indicating that GenAdapt significantly improves DICES w.r.t. the three metrics on the five subjects. For the four other subjects -- \hbox{FULL(5, 3)}, FULL(7,3), FULL(7,4) and FULL(10, 3) -- GenAdapt significantly improves DICES with large or medium effect sizes w.r.t. at least one of these three metrics. For all the subjects, the averages of these three metrics obtained by GenAdapt are better than those obtained by DICES, and there is no case where DICES performs  better than GenAdapt w.r.t. any of these three metrics.

As for the execution time, a larger execution time, as long as it is not worsening the other three metrics, is not a weakness.  For eight subjects, the average execution time of GenAdapt is significantly better than that of DICES and for two, it is the opposite. In those two cases, however, while GenAdapt is slower than DICES on average, it still significantly outperforms DICES w.r.t. the other three metrics. 

\begin{figure}[t]
	\centerline{\includegraphics[width=0.46\columnwidth]{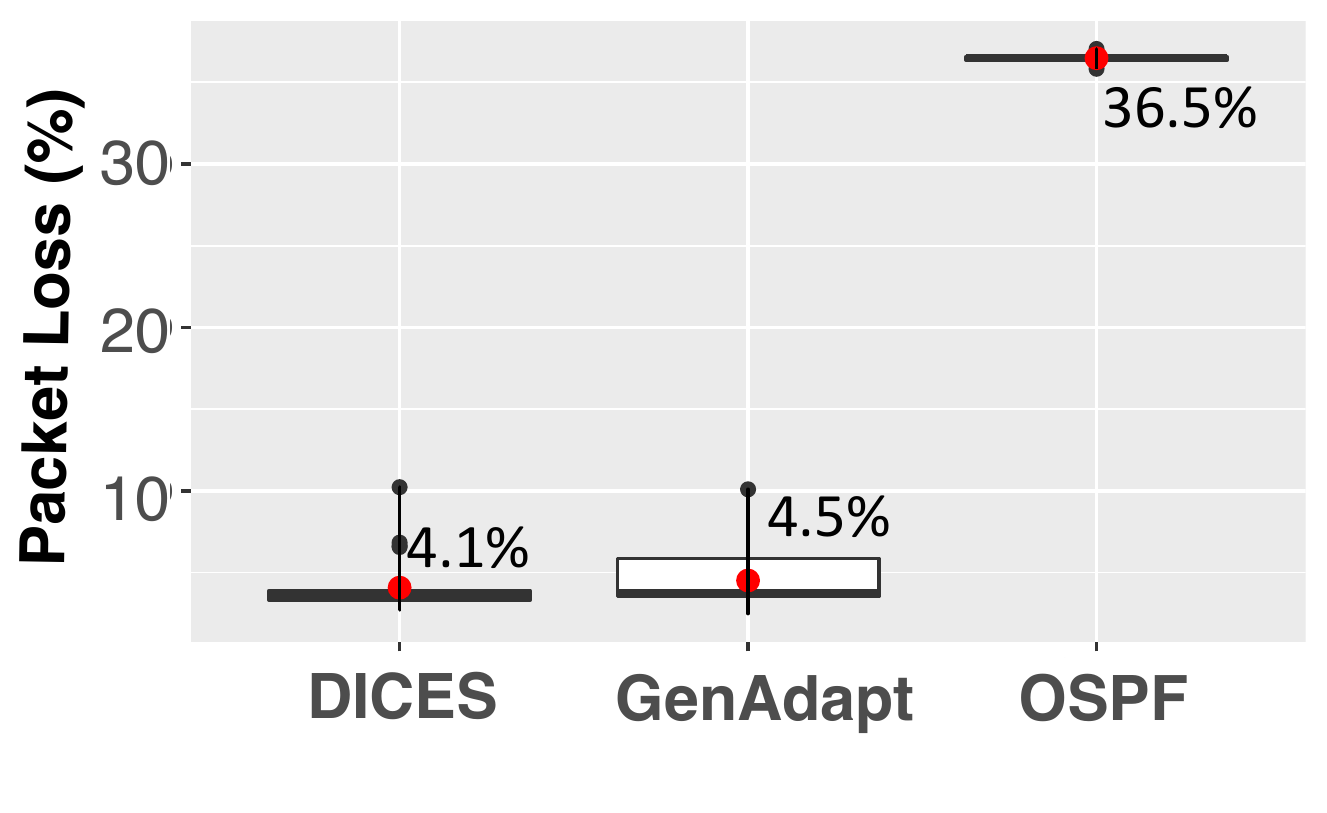}}
	\vspace{-1.2em}
	\caption{Comparing packet-loss values obtained by 30 runs of DICES, GenAdapt and OSPF.}\label{fig:EMSeval}
		\vspace{-.5cm}
\end{figure}

 Figure~\ref{fig:EMSeval} shows the packet-loss values obtained by 30 runs of GenAdapt, DICES and OSPF applied to the industry subject (EMS). All  30 runs of GenAdapt and DICES could resolve the congestion in EMS within the simulation time. However, OSPF incurs high packet loss (avg. $36.5$\%), since its heuristic link weights cannot  prevent congestion in EMS. Both GenAdapt and DICES start with high packet loss, but since they can resolve congestion, packet loss drops quickly, yielding  low averages over the duration of simulation. The differences between the packet-loss values of  GenAdapt and DICES are neither statistically significant ($p$-value = 0.44) nor practically significant -- a packet-loss difference of 0.4\%  is negligible in practice.  The results of Figure~\ref{fig:EMSeval} signify the need for runtime adaptation in EMS. Further, the results in both Table~\ref{tab:GpvsDices} and Figure~\ref{fig:EMSeval}  show that while GenAdapt can reduce congestion occurrences over time compared to DICES, doing so does not come at the cost of being less effective in resolving a one-off congestion in EMS caused by several requests arriving almost at once.

\vspace*{.15cm}
 \resq{The answer to \textbf{RQ1} is that GenAdapt successfully resolves congestions in all the subjects. Compared to DICES, GenAdapt reduces the average number of congestions for six subjects with a high statistical significance, while DICES never outperforms GenAdapt in any of the congestion resolution metrics. Further, for our industry subject, GenAdapt significantly outperforms the standard SDN data-forwarding algorithm, OSPF, in reducing packet loss.}

%% file: table1.tex
\begin{table*}[th]
	\caption{Statistical-test results comparing the average number of congestion occurrences, the duration of congestion, the execution-time and the packet-loss values obtained by 30 runs of GenAdapt versus DICES for our ten synthetic subjects.}
	\label{tab:GpvsDices}
	\vspace*{-.3cm}\hspace*{-.5em}
	\centering
	\scalebox{0.55}{
	\begin{tabular}{c|c|c|c||c|c|c||c|c|c||c|c|c||c|c|c||}
	\toprule
	  & \multicolumn{3}{c||}{ \textbf{\emph{FULL(5,3)}}} & 
	  \multicolumn{3}{c||}{ \textbf{\emph{FULL(5,4)}}} &  
	  \multicolumn{3}{c||}{ \textbf{\emph{FULL(6,3)}} -- {\color{red}\textbf{40\% of DICES runs failed}}} & 
	  \multicolumn{3}{c||}{ \textbf{\emph{FULL(6,4)}} -- {\color{red}\textbf{
	  33.3\% of DICES runs failed}}} & 
	  \multicolumn{3}{c||}{ \textbf{\emph{FULL(7,3)}} -- {\color{red}\textbf{66.6\% of DICES runs failed}}} \\
	 \hline
 & \textbf{avg(G-D)$^*$}& \textbf{p-value} & \textbf{$\bf{\hat{A}_{12}}$} &
 \textbf{avg(G-D)}& \textbf{p-value} & \textbf{$\bf{\hat{A}_{12}}$} &
 \textbf{avg(G-D)}& \textbf{p-value} & \textbf{$\bf{\hat{A}_{12}}$} &
 \textbf{avg(G-D)}& \textbf{p-value} & \textbf{$\bf{\hat{A}_{12}}$} &
 \textbf{avg(G-D)}& \textbf{p-value} & \textbf{$\bf{\hat{A}_{12}}$} \\
 \hline
\# Congestion & 3.1 -- 3.2 & 0.98 & 0.5(N) & 
3.8 -- 4.4 & 0.193 & 0.59(S) & 
3.5 -- 4.5 & 0.003 & 0.74(L) & 
3.8 -- 5.4 & 5.39E-06 & 0.87(L) &
4.1 -- 6 & 9.70E-08 & 1(L) \\

Congestion Duration (s) & 6.2 -- 7.7 & 0.0919  & 0.62(S)& 
9.53 -- 12 & 0.145  & 0.61(S) & 
7.93 -- 12 & 1.69E-06  & 0.86(L) & 
10.03 -- 14.47 & 5.98E-05  & 0.8(L) & 
8.03 -- 16.63 & 1.35E-11 & 1(L) \\

Packet Loss (\%) & 32.1 -- 32.33 & 0.001 & 0.75(L) & 
24.86 -- 24.99 & 0.535 & 0.45(N) & 
24.94 -- 31.73 & 7.96E-11 & 0.94(L) & 
28.4 -- 31.64 & 1.53E-05 & 0.83(L) & 
16.6 -- 19.7 & 0.145 & 0.61(S) \\ 

Exec Time (ms)  & 140.86 -- 392.96 & <2.2E-16 &0.98(L) & 
271.95 -- 440.87 & <2.2E-16 & 0.87(L) & 
302.21 -- 701.06 & <2.2E-16 & 0.9(L) & 
956.47 -- 802.48 & <2.2E-16 & 0.85(L) & 
295.94 -- 949.2 & <2.2E-16 & 0.97(L) \\
\bottomrule
  & \multicolumn{3}{c||}{ \textbf{\emph{FULL(7,4)}}} & 
	  \multicolumn{3}{c||}{ \textbf{\emph{FULL(10,3)}}} &  
	  \multicolumn{3}{c||}{ \textbf{\emph{FULL(10,4)}} -- {\color{red} \textbf{ 10\% of DICES runs failed}}} & 
	  \multicolumn{3}{c||}{ \textbf{\emph{MNP(5,2)}}} & 
	  \multicolumn{3}{c||}{ \textbf{\emph{MNP(8,2)}}} 
	  \\
\hline
& \textbf{avg(G-D)}& \textbf{p-value} & \textbf{$\bf{\hat{A}_{12}}$} &
 \textbf{avg(G-D)}& \textbf{p-value} & \textbf{$\bf{\hat{A}_{12}}$} &
 \textbf{avg(G-D)}& \textbf{p-value} & \textbf{$\bf{\hat{A}_{12}}$} &
 \textbf{avg(G-D)}& \textbf{p-value} & \textbf{$\bf{\hat{A}_{12}}$} &
 \textbf{avg(G-D)}& \textbf{p-value} & \textbf{$\bf{\hat{A}_{12}}$} \\  
\hline
\# Congestion & 2.9 -- 3.1 & 0.063 & 0.59(S) & 
3.7 -- 4 & 0.464 & 0.58(S) & 
4 -- 5.9 & 9.21E-11 & 0.96(L) & 
1.8 -- 2.2 & 0.002 & 0.68(M) & 
2.1 -- 3.4 & 1.55E-08 & 0.9(L) \\

Congestion Duration (s) & 5 -- 5.97 & 0.0157  & 0.68(M) & 7.5 -- 8.47 & 0.029  & 0.65(S) & 8.9 -- 16.23 & 9.53E-11  & 0.98(L) & 3.87 -- 5.23 & 0.0029 & 0.71(M) & 5.57 -- 7.6& 0.0002  & 0.77(L)\\ 

Packet Loss (\%)  & 10.13 -- 19.31 & 8.08E-07 & 0.87(L) & 8.75 -- 13.02 & 0.006 & 0.71(M) & 30.17 -- 34.41 & 1.61E-10 & 0.98(L) & 32.39 -- 32.65 & 0.004 & 0.72(M) & 32.6 -- 33.6 & 3.43E-06 & 0.85(L) \\

Exec Time (ms) & 226.40 -- 576.05 & <2.2E-16 & 0.99(L) & 564.88 -- 814.8 & 4.65E-16 & 0.81(L) & 596.31 -- 1365.28 & <2.2E-16 & 0.86(L) & 381.07 -- 469.13 & 0.099 & 0.59(S) & 787.06 -- 392.62 & <2.2E-16 & 0.08(L)\\
\bottomrule
	\end{tabular}}

\vspace*{.1cm}	
{\scriptsize 
\hspace{.5cm} * The avg(G-D) column shows the average (avg) metrics for GenAdapt (G) versus those for DICES (D).\hfill\mbox{}}
\vspace*{-.35cm}
\end{table*}

%% file: Rq2.tex
\subsection{RQ2 -- Scalability}
To answer RQ2, we perform two sets of studies. In the first set, we increase the network size and in the second set -- the number of requests. Specifically, first, we create ten synthetic networks with the FULL topology and having  5, 10, \ldots , 50 nodes; and, for each network, we generate five requests with the same bandwidth at once to reach a total bandwidth of 150Mbps.  Second, we create a five-node network  with the FULL topology and execute ten different experiments by subsequently generating  5, 10, \ldots , 50 requests at once such that for each experiment, the requests have the same bandwidth and the sum of the requests' bandwidths is 150Mbps. For both experimental sets, we record the execution time of the configuration of GenAdapt used for the FULL topology as described in Section~\ref{sec:rq1}. Our experimental setup in RQ2 follows that used by DICES for scalability analysis. We focus on the FULL topology for scalability analysis because networks with this topology have considerably more links and pose a bigger challenge for self-adaptation planning, as evidenced by the failure of  DICES over several networks with the FULL topology (see Table~\ref{tab:GpvsDices}).

\begin{figure}[t]
	\centerline{\includegraphics[width=\columnwidth]{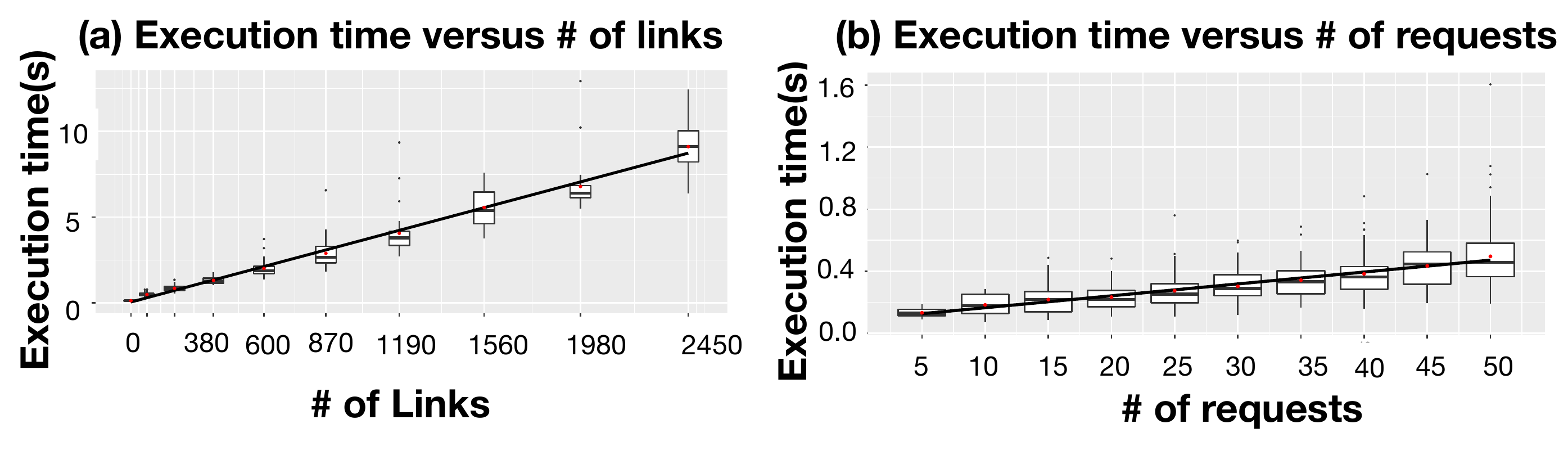}}
	\vspace{-1.2em}
	\caption{Execution times of GenAdapt versus (a) number  of links and (b) number of requests.}
	\label{fig:rq2}
		\vspace{-.8cm}
\end{figure}

Figure~\ref{fig:rq2} shows the execution time of GenAdapt versus the number of network links (Figures~\ref{fig:rq2}(a)), and  versus the number of requests (Figure~\ref{fig:rq2}(b)). For each network size and for each number of requests, GenAdapt is executed 30 times. For each diagram in Figures~\ref{fig:rq2}, we have fitted a linear regression line ($\mathit{time} = -0.019 + 0.0036 \times \mathit{links}$ in Figure~\ref{fig:rq2}(a); and $\mathit{time} = 0.087  + 0.0077 \times \mathit{links}$ in Figure~\ref{fig:rq2}(b)). In both cases, we obtain a $p$-value of $< 2.2e-16$, indicating that the models fit the data well.

\vspace*{.15cm}
\resq{The answer to \textbf{RQ2} is that the execution time of GenAdapt is linear in the number of requests and in the size of the network.}
\vspace*{-.15cm}

%% file: threats.tex
\subsection{Threats to Validity}
Construct and external validity are the validity aspects most pertinent to our  evaluation.

\textbf{Construct validity.} Our experiments are simulation-based; the extent to which the measurements obtained through simulation are reflective of the real world is therefore an important factor to consider. To this end, we note that the tools that our simulator builds on, i.e., ONOS and Mininet, are widely considered to be \emph{high-fidelity} network simulators~\cite{DBLP:journals/csur/ZhuKSXLDG21,DBLP:conf/cnsm/AzzouniBP17,Berde:14,Bianco:17}.  While the high fidelity of our simulator provides confidence about our results, future experimentation with real IoT networks remains necessary.


\textbf{External validity.} We scoped our experiments to two network topologies only. As discussed in Section~\ref{sec:rq1}, the two topologies are the two extreme ends of the spectrum in terms of path overlaps between node pairs -- the main determinant of complexity for GenAdapt. We expect that, for other topologies lending themselves  to congestion resolution via re-routing, our approach will behave within the same range as seen in our experiments. To ensure adequate coverage, our evaluation examined 18 networks based on the two topologies considered. In addition, our evaluation included a real industrial IoT network with its own topology. The number, size and complexity of the networks in our experiments are comparable to those in the literature, e.g.,~\cite{ShinNSB0Z20,DBLP:conf/cnsm/AzzouniBP17,DBLP:conf/cscwd/LiuZGQ18,DBLP:journals/csur/ZhuKSXLDG21,Agarwal:13}. A second consideration related to external validity is the tuning of our approach. The only parameter of our approach that requires tuning by the user is the number of generations of GenAdapt. As discussed in Section~\ref{sec:rq1}, tuning this parameter is straightforward based on the easily measurable objective of keeping the execution time of GenAdapt less than $\delta$ (i.e., the time interval between consecutive invocations of GenAdapt). In practice, engineers can use our simulator to tune this parameter for their specific network topology and traffic profile.

%% file: related.tex

\section{Related Work}
\label{sec:related}
We compare our approach with related work in software engineering for self-adaptive systems and in congestion control for SDN. 




\textbf{Self-adaptive systems.} Engineering self-adaptive systems, including the principles underlying the construction, maintenance and evolution of such systems, have been studied from different angles and for different domains~\cite{Krupitzer:18,Coker:15,Bhuiyan:17,Anaya:14,Stein:16,DBLP:books/daglib/p/GarlanSC09,ChengRM13}.  Our work relates to dynamic adaptive search-based software engineering~\cite{Harman:12} which uses a blend of artificial intelligence (AI) and optimization to adapt system properties. Prior research in this field has used search algorithms for different purposes such as configuring properties of self-adaptive systems~\cite{Ramirez:09,Zoghi:16} and improving the design and architecture of such systems~\cite{Menasce:11,Andrade:13,Ramirez:10}. The closest work to ours is DICES~\cite{ShinNSB0Z20}, which is used  as a baseline for our experiments. Our approach, in contrast to DICES,  is generative and aims to reduce the need for invoking adaptations over time. 

Our work relates to AI-enabled adaptation~\cite{DBLP:conf/icse/QuinWBBM19,DBLP:conf/seams/GheibiWQ21}. Among different AI methods, we use GP because of its flexibility to learn formulas that match our specific grammar for link-weight functions. To our knowledge, we are the first to apply GP for adapting IoT networks and reducing the need for frequent \hbox{adaptations in this domain.}

\textbf{Congestion control in SDN.} Network congestion resolution has been studied extensively~\cite{Mathis:96,Alizadeh:10,He:16,Betzler:16}.
Some congestion resolution approaches work by adjusting data transmission rates~\cite{Betzler:16}. Several others focus on traffic  engineering to provide solutions for better traffic control, better traffic operation, and better traffic management, e.g., by multi-path routing~\cite{DBLP:journals/wicomm/HanDGSW15},   creating a new routing technique~\cite{DBLP:conf/globecom/MaoFTKAIM17},  and flow-based routing~\cite{DBLP:conf/globecom/AmokraneLBP13}. Our work is related to the research that utilizes the additional flexibility offered by the programmable control in the SDN architecture~\cite{Brandt:16,Hong:13,Chiang:18,Agarwal:13,Gay:17,Huang:16,ShinNSB0Z20,Yashar:12}. Within this line of research, congestion control is commonly cast as an optimization problem~\cite{Chiang:18}, to be solved using local search~\cite{Gay:17} or linear programming~\cite{Agarwal:13}.
Some techniques~\cite{Yashar:12} adapt the network parameters at runtime using a pre-defined set of rules. In contrast to our work, none of these techniques evolve the rules in response to feedback from SDN monitoring.

%% file: conclusions.tex
\vspace*{-.2cm}
\section{Conclusion}
\label{sec:conclusions}
We presented GenAdapt -- an adaptive approach that uses genetic programming for resolving network congestion in SDN-based IoT~networks. 
While existing self-adaptation research focuses on modifying a running system via producing individual and concrete elements, GenAdapt is generative and modifies the logic of the running system so that the system itself can generate the concrete elements without needing frequent adaptations. We used 18 synthetic and one industrial network to compare GenAdapt against two baseline techniques:  DICES~\cite{ShinNSB0Z20}  and OSPF~\cite{Coltun:08,Cisco:05}. GenAdapt successfully resolved all congestion occurrences in our experimental networks, while DICES failed to do so in four of them. Further, compared to DICES, GenAdapt reduced the number of congestion occurrences and outperformed OSPF  in reducing packet loss. 



%% file: paper.bbl

\begin{thebibliography}{87}


\ifx \showCODEN    \undefined \def \showCODEN     #1{\unskip}     \fi
\ifx \showDOI      \undefined \def \showDOI       #1{#1}\fi
\ifx \showISBNx    \undefined \def \showISBNx     #1{\unskip}     \fi
\ifx \showISBNxiii \undefined \def \showISBNxiii  #1{\unskip}     \fi
\ifx \showISSN     \undefined \def \showISSN      #1{\unskip}     \fi
\ifx \showLCCN     \undefined \def \showLCCN      #1{\unskip}     \fi
\ifx \shownote     \undefined \def \shownote      #1{#1}          \fi
\ifx \showarticletitle \undefined \def \showarticletitle #1{#1}   \fi
\ifx \showURL      \undefined \def \showURL       {\relax}        \fi
\providecommand\bibfield[2]{#2}
\providecommand\bibinfo[2]{#2}
\providecommand\natexlab[1]{#1}
\providecommand\showeprint[2][]{arXiv:#2}

\bibitem[\protect\citeauthoryear{??}{Cis}{2005}]%
        {Cisco:05}
 \bibinfo{year}{2005}\natexlab{}.
\newblock \bibinfo{title}{Cisco. OSPF Design Guide}.
\newblock \bibinfo{howpublished}{Documentation at
  https://www.cisco.com/c/en/us/support/docs/ip/open-shortest-path-first-ospf/7039-1.html}.
\newblock


\bibitem[\protect\citeauthoryear{??}{app}{2021}]%
        {appedix}
 \bibinfo{year}{2021}\natexlab{}.
\newblock \bibinfo{title}{GenAdapt}.
\newblock
\newblock
\urldef\tempurl%
\url{https://figshare.com/s/de6eb6e61816401b5c9e}
\showURL{%
\tempurl}


\bibitem[\protect\citeauthoryear{Agarwal, Kodialam, and Lakshman}{Agarwal
  et~al\mbox{.}}{2013}]%
        {Agarwal:13}
\bibfield{author}{\bibinfo{person}{Sugam Agarwal}, \bibinfo{person}{Murali~S.
  Kodialam}, {and} \bibinfo{person}{T.~V. Lakshman}.}
  \bibinfo{year}{2013}\natexlab{}.
\newblock \showarticletitle{Traffic Engineering in Software Defined Networks}.
  In \bibinfo{booktitle}{\emph{Proceedings of the 2013 Annual IEEE
  International Conference on Computer Communications (INFOCOM'13)}}.
  \bibinfo{pages}{2211--2219}.
\newblock


\bibitem[\protect\citeauthoryear{Akyildiz, Lee, Wang, Luo, and Chou}{Akyildiz
  et~al\mbox{.}}{2014}]%
        {Akyildiz:14}
\bibfield{author}{\bibinfo{person}{Ian~F. Akyildiz}, \bibinfo{person}{Ahyoung
  Lee}, \bibinfo{person}{Pu Wang}, \bibinfo{person}{Min Luo}, {and}
  \bibinfo{person}{Wu Chou}.} \bibinfo{year}{2014}\natexlab{}.
\newblock \showarticletitle{A Roadmap for Traffic Engineering in
  {SDN}-{OpenFlow} Networks}.
\newblock \bibinfo{journal}{\emph{Computer Networks}}  \bibinfo{volume}{71}
  (\bibinfo{year}{2014}), \bibinfo{pages}{1--30}.
\newblock


\bibitem[\protect\citeauthoryear{Alizadeh, Greenberg, Maltz, Padhye, Patel,
  Prabhakar, Sengupta, and Sridharan}{Alizadeh et~al\mbox{.}}{2010}]%
        {Alizadeh:10}
\bibfield{author}{\bibinfo{person}{Mohammad Alizadeh},
  \bibinfo{person}{Albert~G. Greenberg}, \bibinfo{person}{David~A. Maltz},
  \bibinfo{person}{Jitendra Padhye}, \bibinfo{person}{Parveen Patel},
  \bibinfo{person}{Balaji Prabhakar}, \bibinfo{person}{Sudipta Sengupta}, {and}
  \bibinfo{person}{Murari Sridharan}.} \bibinfo{year}{2010}\natexlab{}.
\newblock \showarticletitle{Data Center {TCP} {(DCTCP)}}. In
  \bibinfo{booktitle}{\emph{Proceedings of the 2010 ACM Conference on Special
  Interest Group on Data Communication (SIGCOMM'10)}}. \bibinfo{pages}{63--74}.
\newblock


\bibitem[\protect\citeauthoryear{Alrajeh, Cailliau, and van Lamsweerde}{Alrajeh
  et~al\mbox{.}}{2020}]%
        {DBLP:conf/icse/AlrajehCL20}
\bibfield{author}{\bibinfo{person}{Dalal Alrajeh}, \bibinfo{person}{Antoine
  Cailliau}, {and} \bibinfo{person}{Axel van Lamsweerde}.}
  \bibinfo{year}{2020}\natexlab{}.
\newblock \showarticletitle{Adapting requirements models to varying
  environments}. In \bibinfo{booktitle}{\emph{{ICSE} '20: 42nd International
  Conference on Software Engineering, Seoul, South Korea, 27 June - 19 July,
  2020}}, \bibfield{editor}{\bibinfo{person}{Gregg Rothermel} {and}
  \bibinfo{person}{Doo{-}Hwan Bae}} (Eds.). \bibinfo{publisher}{{ACM}},
  \bibinfo{pages}{50--61}.
\newblock


\bibitem[\protect\citeauthoryear{Amin, Reisslein, and Shah}{Amin
  et~al\mbox{.}}{2018}]%
        {Amin:18}
\bibfield{author}{\bibinfo{person}{Rashid Amin}, \bibinfo{person}{Martin
  Reisslein}, {and} \bibinfo{person}{Nadir Shah}.}
  \bibinfo{year}{2018}\natexlab{}.
\newblock \showarticletitle{Hybrid {SDN} Networks: {A} Survey of Existing
  Approaches}.
\newblock \bibinfo{journal}{\emph{{IEEE} Communications Surveys and Tutorials}}
  \bibinfo{volume}{20}, \bibinfo{number}{4} (\bibinfo{year}{2018}),
  \bibinfo{pages}{3259--3306}.
\newblock


\bibitem[\protect\citeauthoryear{Amokrane, Langar, Boutaba, and
  Pujolle}{Amokrane et~al\mbox{.}}{2013}]%
        {DBLP:conf/globecom/AmokraneLBP13}
\bibfield{author}{\bibinfo{person}{Ahmed Amokrane}, \bibinfo{person}{Rami
  Langar}, \bibinfo{person}{Raouf Boutaba}, {and} \bibinfo{person}{Guy
  Pujolle}.} \bibinfo{year}{2013}\natexlab{}.
\newblock \showarticletitle{Online flow-based energy efficient management in
  Wireless Mesh Networks}. In \bibinfo{booktitle}{\emph{2013 {IEEE} Global
  Communications Conference, {GLOBECOM} 2013, Atlanta, GA, USA, December 9-13,
  2013}}. \bibinfo{publisher}{{IEEE}}, \bibinfo{pages}{329--335}.
\newblock


\bibitem[\protect\citeauthoryear{Anaya, Simko, Bourcier, Plouzeau, and
  J{\'{e}}z{\'{e}}quel}{Anaya et~al\mbox{.}}{2014}]%
        {Anaya:14}
\bibfield{author}{\bibinfo{person}{Ivan Dario~Paez Anaya},
  \bibinfo{person}{Viliam Simko}, \bibinfo{person}{Johann Bourcier},
  \bibinfo{person}{No{\"{e}}l Plouzeau}, {and} \bibinfo{person}{Jean{-}Marc
  J{\'{e}}z{\'{e}}quel}.} \bibinfo{year}{2014}\natexlab{}.
\newblock \showarticletitle{A Prediction-driven Adaptation Approach for
  Self-Adaptive Sensor Networks}. In \bibinfo{booktitle}{\emph{Proceedings of
  the 9th International Symposium on Software Engineering for Adaptive and
  Self-Managing Systems {SEAMS'14}}}. \bibinfo{pages}{145--154}.
\newblock


\bibitem[\protect\citeauthoryear{Andrade and de~A.~Mac\^{e}do}{Andrade and
  de~A.~Mac\^{e}do}{2013}]%
        {Andrade:13}
\bibfield{author}{\bibinfo{person}{Sandro~S. Andrade} {and}
  \bibinfo{person}{Raimundo~Jos\'{e} de A.~Mac\^{e}do}.}
  \bibinfo{year}{2013}\natexlab{}.
\newblock \showarticletitle{A Search-Based Approach for Architectural Design of
  Feedback Control Concerns in Self-Adaptive Systems}. In
  \bibinfo{booktitle}{\emph{Proceedings of the 7th IEEE International
  Conference on Self-Adaptive and Self-Organizing Systems (SASO'13)}}.
  \bibinfo{pages}{61--70}.
\newblock


\bibitem[\protect\citeauthoryear{Apostolopoulos, Kamat, Williams, Gu{\'{e}}rin,
  Orda, and Przygienda}{Apostolopoulos et~al\mbox{.}}{1999}]%
        {Apostolopoulos:99}
\bibfield{author}{\bibinfo{person}{G. Apostolopoulos}, \bibinfo{person}{S.
  Kamat}, \bibinfo{person}{D. Williams}, \bibinfo{person}{R. Gu{\'{e}}rin},
  \bibinfo{person}{A. Orda}, {and} \bibinfo{person}{T. Przygienda}.}
  \bibinfo{year}{1999}\natexlab{}.
\newblock \showarticletitle{QoS Routing Mechanisms and {OSPF} Extensions}.
\newblock \bibinfo{journal}{\emph{{RFC}}}  \bibinfo{volume}{2676}
  (\bibinfo{year}{1999}), \bibinfo{pages}{1--50}.
\newblock


\bibitem[\protect\citeauthoryear{Azzouni, Boutaba, and Pujolle}{Azzouni
  et~al\mbox{.}}{2017}]%
        {DBLP:conf/cnsm/AzzouniBP17}
\bibfield{author}{\bibinfo{person}{Abdelhadi Azzouni}, \bibinfo{person}{Raouf
  Boutaba}, {and} \bibinfo{person}{Guy Pujolle}.}
  \bibinfo{year}{2017}\natexlab{}.
\newblock \showarticletitle{NeuRoute: Predictive dynamic routing for
  software-defined networks}. In \bibinfo{booktitle}{\emph{13th International
  Conference on Network and Service Management, {CNSM} 2017, Tokyo, Japan,
  November 26-30, 2017}}. \bibinfo{publisher}{{IEEE} Computer Society},
  \bibinfo{pages}{1--6}.
\newblock


\bibitem[\protect\citeauthoryear{Berde, Gerola, Hart, Higuchi, Kobayashi,
  Koide, O'Connor, Radoslavov, Snow, and Parulkar}{Berde et~al\mbox{.}}{2014}]%
        {Berde:14}
\bibfield{author}{\bibinfo{person}{Pankaj Berde}, \bibinfo{person}{Matteo
  Gerola}, \bibinfo{person}{Jonathan Hart}, \bibinfo{person}{Yuta Higuchi},
  \bibinfo{person}{Masayoshi Kobayashi}, \bibinfo{person}{Toshio Koide},
  \bibinfo{person}{Bob Lantzand~Brian O'Connor}, \bibinfo{person}{Pavlin
  Radoslavov}, \bibinfo{person}{William Snow}, {and} \bibinfo{person}{Guru
  Parulkar}.} \bibinfo{year}{2014}\natexlab{}.
\newblock \showarticletitle{{ONOS}: Towards an Open, Distributed {SDN} {OS}}.
  In \bibinfo{booktitle}{\emph{Proceedings of the 3rd Workshop on Hot Topics in
  Software Defined Networking (HotSDN'14)}}. \bibinfo{pages}{1--6}.
\newblock


\bibitem[\protect\citeauthoryear{Betzler, Gomez, Demirkol, and
  Paradells}{Betzler et~al\mbox{.}}{2016}]%
        {Betzler:16}
\bibfield{author}{\bibinfo{person}{August Betzler}, \bibinfo{person}{Carles
  Gomez}, \bibinfo{person}{Ilker Demirkol}, {and} \bibinfo{person}{Josep
  Paradells}.} \bibinfo{year}{2016}\natexlab{}.
\newblock \showarticletitle{{CoAP} Congestion Control for the Internet of
  Things}.
\newblock \bibinfo{journal}{\emph{{IEEE} Communications Magazine}}
  \bibinfo{volume}{54}, \bibinfo{number}{7} (\bibinfo{year}{2016}),
  \bibinfo{pages}{154--160}.
\newblock


\bibitem[\protect\citeauthoryear{Bhuiyan, Wu, Wang, Wang, and Hassan}{Bhuiyan
  et~al\mbox{.}}{2017}]%
        {Bhuiyan:17}
\bibfield{author}{\bibinfo{person}{Md~Zakirul~Alam Bhuiyan},
  \bibinfo{person}{Jie Wu}, \bibinfo{person}{Guojun Wang},
  \bibinfo{person}{Tian Wang}, {and} \bibinfo{person}{Mohammad~Mehedi Hassan}.}
  \bibinfo{year}{2017}\natexlab{}.
\newblock \showarticletitle{{e-Sampling}: Event-Sensitive Autonomous Adaptive
  Sensing and Low-Cost Monitoring in Networked Sensing Systems}.
\newblock \bibinfo{journal}{\emph{ACM Transactions on Autonomous and Adaptive
  Systems (TAAS)}} \bibinfo{volume}{12}, \bibinfo{number}{1}
  (\bibinfo{year}{2017}), \bibinfo{pages}{1:1--1:29}.
\newblock


\bibitem[\protect\citeauthoryear{Bianco, Giaccone, Mahmood, Ullio, and
  Vercellone}{Bianco et~al\mbox{.}}{2015}]%
        {Bianco:15}
\bibfield{author}{\bibinfo{person}{Andrea Bianco}, \bibinfo{person}{Paolo
  Giaccone}, \bibinfo{person}{Ahsan Mahmood}, \bibinfo{person}{Mario Ullio},
  {and} \bibinfo{person}{Vinicio Vercellone}.} \bibinfo{year}{2015}\natexlab{}.
\newblock \showarticletitle{Evaluating the {SDN} control traffic in large {ISP}
  networks}. In \bibinfo{booktitle}{\emph{Proceedings of the 2015 {IEEE}
  International Conference on Communications (ICC'15)}}.
  \bibinfo{pages}{5248--5253}.
\newblock


\bibitem[\protect\citeauthoryear{Bianco, Giaccone, Mashayekhi, Ullio, and
  Vercellone}{Bianco et~al\mbox{.}}{2017}]%
        {Bianco:17}
\bibfield{author}{\bibinfo{person}{Andrea Bianco}, \bibinfo{person}{Paolo
  Giaccone}, \bibinfo{person}{Reza Mashayekhi}, \bibinfo{person}{Mario Ullio},
  {and} \bibinfo{person}{Vinicio Vercellone}.} \bibinfo{year}{2017}\natexlab{}.
\newblock \showarticletitle{Scalability of {ONOS} reactive forwarding
  applications in {ISP} networks}.
\newblock \bibinfo{journal}{\emph{Computer Communications}}
  \bibinfo{volume}{102} (\bibinfo{year}{2017}), \bibinfo{pages}{130--138}.
\newblock


\bibitem[\protect\citeauthoryear{Borg, Abdessalem, Nejati, Jegeden, and
  Shin}{Borg et~al\mbox{.}}{2021}]%
        {paper2}
\bibfield{author}{\bibinfo{person}{Markus Borg}, \bibinfo{person}{Raja~Ben
  Abdessalem}, \bibinfo{person}{Shiva Nejati},
  \bibinfo{person}{Fran{\c{c}}ois{-}Xavier Jegeden}, {and}
  \bibinfo{person}{Donghwan Shin}.} \bibinfo{year}{2021}\natexlab{}.
\newblock \showarticletitle{Digital Twins Are Not Monozygotic -
  Cross-Replicating {ADAS} Testing in Two Industry-Grade Automotive
  Simulators}. In \bibinfo{booktitle}{\emph{14th {IEEE} Conference on Software
  Testing, Verification and Validation, {ICST} 2021}}.
  \bibinfo{publisher}{{IEEE}}, \bibinfo{pages}{383--393}.
\newblock


\bibitem[\protect\citeauthoryear{Botta, Dainotti, and Pescap\`{e}}{Botta
  et~al\mbox{.}}{2012}]%
        {Botta:12}
\bibfield{author}{\bibinfo{person}{Alessio Botta}, \bibinfo{person}{Alberto
  Dainotti}, {and} \bibinfo{person}{Antonio Pescap\`{e}}.}
  \bibinfo{year}{2012}\natexlab{}.
\newblock \showarticletitle{A Tool for The Generation of Realistic Network
  Workload for Emerging Networking Scenarios}.
\newblock \bibinfo{journal}{\emph{Computer Networks}} \bibinfo{volume}{56},
  \bibinfo{number}{15} (\bibinfo{year}{2012}), \bibinfo{pages}{3531--3547}.
\newblock


\bibitem[\protect\citeauthoryear{Brandt, Foerster, and Wattenhofer}{Brandt
  et~al\mbox{.}}{2016}]%
        {Brandt:16}
\bibfield{author}{\bibinfo{person}{Sebastian Brandt},
  \bibinfo{person}{{Klaus-Tycho} Foerster}, {and} \bibinfo{person}{Roger
  Wattenhofer}.} \bibinfo{year}{2016}\natexlab{}.
\newblock \showarticletitle{On Consistent Migration of Flows in {SDN}s}. In
  \bibinfo{booktitle}{\emph{Proceedings of the 2016 Annual IEEE International
  Conference on Computer Communications (INFOCOM'16)}}. \bibinfo{pages}{1--9}.
\newblock


\bibitem[\protect\citeauthoryear{Capon}{Capon}{1991}]%
        {capon:91}
\bibfield{author}{\bibinfo{person}{J.~Anthony Capon}.}
  \bibinfo{year}{1991}\natexlab{}.
\newblock \bibinfo{booktitle}{\emph{Elementary Statistics for the Social
  Sciences: Study Guide}}.
\newblock \bibinfo{publisher}{Wadsworth Publishing Company},
  \bibinfo{address}{Belmont, CA, USA}.
\newblock


\bibitem[\protect\citeauthoryear{Caria, Das, and Jukan}{Caria
  et~al\mbox{.}}{2015}]%
        {Caria:15}
\bibfield{author}{\bibinfo{person}{Marcel Caria}, \bibinfo{person}{Tamal Das},
  {and} \bibinfo{person}{Admela Jukan}.} \bibinfo{year}{2015}\natexlab{}.
\newblock \showarticletitle{Divide and conquer: Partitioning {OSPF} networks
  with {SDN}}. In \bibinfo{booktitle}{\emph{Proceedings of the 2015 {IFIP/IEEE}
  International Symposium on Integrated Network Management (IM'15)}}.
  \bibinfo{pages}{467--474}.
\newblock


\bibitem[\protect\citeauthoryear{Cheng, Ramirez, and McKinley}{Cheng
  et~al\mbox{.}}{2013}]%
        {ChengRM13}
\bibfield{author}{\bibinfo{person}{Betty H.~C. Cheng},
  \bibinfo{person}{Andres~J. Ramirez}, {and} \bibinfo{person}{Philip~K.
  McKinley}.} \bibinfo{year}{2013}\natexlab{}.
\newblock \showarticletitle{Harnessing evolutionary computation to enable
  dynamically adaptive systems to manage uncertainty}. In
  \bibinfo{booktitle}{\emph{1st International Workshop on Combining Modelling
  and Search-Based Software Engineering, CMSBSE@ICSE 2013, San Francisco, CA,
  USA, May 20, 2013}}. \bibinfo{publisher}{{IEEE} Computer Society},
  \bibinfo{pages}{1--6}.
\newblock


\bibitem[\protect\citeauthoryear{Chiang, Kuo, Shen, Yang, and Chen}{Chiang
  et~al\mbox{.}}{2018}]%
        {Chiang:18}
\bibfield{author}{\bibinfo{person}{Sheng{-}Hao Chiang},
  \bibinfo{person}{Jian{-}Jhih Kuo}, \bibinfo{person}{Shan{-}Hsiang Shen},
  \bibinfo{person}{De{-}Nian Yang}, {and} \bibinfo{person}{Wen{-}Tsuen Chen}.}
  \bibinfo{year}{2018}\natexlab{}.
\newblock \showarticletitle{Online Multicast Traffic Engineering for
  Software-Defined Networks}. In \bibinfo{booktitle}{\emph{Proceedings of the
  2018 Annual IEEE International Conference on Computer Communications
  (INFOCOM'18)}}. \bibinfo{pages}{414--422}.
\newblock


\bibitem[\protect\citeauthoryear{Coker, Garlan, and Goues}{Coker
  et~al\mbox{.}}{2015}]%
        {Coker:15}
\bibfield{author}{\bibinfo{person}{Zack Coker}, \bibinfo{person}{David Garlan},
  {and} \bibinfo{person}{Claire~Le Goues}.} \bibinfo{year}{2015}\natexlab{}.
\newblock \showarticletitle{{SASS}: Self-adaptation Using Stochastic Search}.
  In \bibinfo{booktitle}{\emph{Proceedings of the 10th International Symposium
  on Software Engineering for Adaptive and Self-Managing Systems {SEAMS'15}}}.
  \bibinfo{pages}{168--174}.
\newblock


\bibitem[\protect\citeauthoryear{Coltun, Ferguson, Moy, and Lindem}{Coltun
  et~al\mbox{.}}{2008}]%
        {Coltun:08}
\bibfield{author}{\bibinfo{person}{Rob Coltun}, \bibinfo{person}{Dennis
  Ferguson}, \bibinfo{person}{John Moy}, {and} \bibinfo{person}{Acee Lindem}.}
  \bibinfo{year}{2008}\natexlab{}.
\newblock \bibinfo{booktitle}{\emph{{OSPF} for IPv6}}.
\newblock \bibinfo{type}{Internet Standard} {RFC} 5340.
  \bibinfo{institution}{Network Working Group}.
\newblock


\bibitem[\protect\citeauthoryear{Cormen, Leiserson, Rivest, and Stein}{Cormen
  et~al\mbox{.}}{2009}]%
        {Cormen:09}
\bibfield{author}{\bibinfo{person}{Thomas~H. Cormen},
  \bibinfo{person}{Charles~E. Leiserson}, \bibinfo{person}{Ronald~L. Rivest},
  {and} \bibinfo{person}{Clifford Stein}.} \bibinfo{year}{2009}\natexlab{}.
\newblock \bibinfo{booktitle}{\emph{Introduction to Algorithms}
  (\bibinfo{edition}{3rd} ed.)}.
\newblock \bibinfo{publisher}{The MIT Press}.
\newblock


\bibitem[\protect\citeauthoryear{DeVries and Cheng}{DeVries and Cheng}{2017}]%
        {DBLP:conf/models/DeVriesC17}
\bibfield{author}{\bibinfo{person}{Byron DeVries} {and} \bibinfo{person}{Betty
  H.~C. Cheng}.} \bibinfo{year}{2017}\natexlab{}.
\newblock \showarticletitle{Using Models at Run Time to Detect Incomplete and
  Inconsistent Requirements}. In \bibinfo{booktitle}{\emph{Proceedings of
  {MODELS} 2017 Satellite Events, Austin, TX, USA, September, 17, 2017}}
  \emph{(\bibinfo{series}{{CEUR} Workshop Proceedings},
  Vol.~\bibinfo{volume}{2019})}. \bibinfo{publisher}{CEUR-WS.org},
  \bibinfo{pages}{201--209}.
\newblock


\bibitem[\protect\citeauthoryear{Feldt and Yoo}{Feldt and Yoo}{2020}]%
        {FeldtY20}
\bibfield{author}{\bibinfo{person}{Robert Feldt} {and} \bibinfo{person}{Shin
  Yoo}.} \bibinfo{year}{2020}\natexlab{}.
\newblock \showarticletitle{Flexible Probabilistic Modeling for Search Based
  Test Data Generation}. In \bibinfo{booktitle}{\emph{{ICSE} '20: 42nd
  International Conference on Software Engineering, SBST Workshop, Seoul,
  Republic of Korea, 27 June - 19 July, 2020}}. \bibinfo{publisher}{{ACM}},
  \bibinfo{pages}{537--540}.
\newblock


\bibitem[\protect\citeauthoryear{Filieri, Hoffmann, and Maggio}{Filieri
  et~al\mbox{.}}{2015}]%
        {Filieri:15}
\bibfield{author}{\bibinfo{person}{Antonio Filieri}, \bibinfo{person}{Henry
  Hoffmann}, {and} \bibinfo{person}{Martina Maggio}.}
  \bibinfo{year}{2015}\natexlab{}.
\newblock \showarticletitle{Automated Design of Self-Adaptive Software with
  Control-Theoretical Formal Guarantees}. In \bibinfo{booktitle}{\emph{Software
  Engineering {\&} Management 2015, Multikonferenz der GI-Fachbereiche
  Softwaretechnik {(SWT)} und Wirtschaftsinformatik (WI), {FA} WI-MAW, 17.
  M{\"{a}}rz - 20. M{\"{a}}rz 2015, Dresden, Germany}}
  \emph{(\bibinfo{series}{{LNI}}, Vol.~\bibinfo{volume}{{P-239}})},
  \bibfield{editor}{\bibinfo{person}{Uwe A{\ss}mann}, \bibinfo{person}{Birgit
  Demuth}, \bibinfo{person}{Thorsten Spitta}, \bibinfo{person}{Georg
  P{\"{u}}schel}, {and} \bibinfo{person}{Ronny Kaiser}} (Eds.).
  \bibinfo{pages}{112--113}.
\newblock


\bibitem[\protect\citeauthoryear{Fortz and Thorup}{Fortz and Thorup}{2000}]%
        {FortzT00}
\bibfield{author}{\bibinfo{person}{Bernard Fortz} {and} \bibinfo{person}{Mikkel
  Thorup}.} \bibinfo{year}{2000}\natexlab{}.
\newblock \showarticletitle{Internet Traffic Engineering by Optimizing {OSPF}
  Weights}. In \bibinfo{booktitle}{\emph{Proceedings {IEEE} {INFOCOM} 2000, The
  Conference on Computer Communications, Nineteenth Annual Joint Conference of
  the {IEEE} Computer and Communications Societies, Reaching the Promised Land
  of Communications, Tel Aviv, Israel, March 26-30, 2000}}.
  \bibinfo{publisher}{{IEEE} Computer Society}, \bibinfo{pages}{519--528}.
\newblock


\bibitem[\protect\citeauthoryear{Fortz and Thorup}{Fortz and Thorup}{2002}]%
        {FortzT02}
\bibfield{author}{\bibinfo{person}{Bernard Fortz} {and} \bibinfo{person}{Mikkel
  Thorup}.} \bibinfo{year}{2002}\natexlab{}.
\newblock \showarticletitle{Optimizing {OSPF/IS-IS} weights in a changing
  world}.
\newblock \bibinfo{journal}{\emph{{IEEE} J. Sel. Areas Commun.}}
  \bibinfo{volume}{20}, \bibinfo{number}{4} (\bibinfo{year}{2002}),
  \bibinfo{pages}{756--767}.
\newblock


\bibitem[\protect\citeauthoryear{Fortz and Thorup}{Fortz and Thorup}{2004}]%
        {FortzT04}
\bibfield{author}{\bibinfo{person}{Bernard Fortz} {and} \bibinfo{person}{Mikkel
  Thorup}.} \bibinfo{year}{2004}\natexlab{}.
\newblock \showarticletitle{Increasing Internet Capacity Using Local Search}.
\newblock \bibinfo{journal}{\emph{Comput. Optim. Appl.}} \bibinfo{volume}{29},
  \bibinfo{number}{1} (\bibinfo{year}{2004}), \bibinfo{pages}{13--48}.
\newblock


\bibitem[\protect\citeauthoryear{Garlan, Schmerl, and Cheng}{Garlan
  et~al\mbox{.}}{2009}]%
        {DBLP:books/daglib/p/GarlanSC09}
\bibfield{author}{\bibinfo{person}{David Garlan}, \bibinfo{person}{Bradley~R.
  Schmerl}, {and} \bibinfo{person}{Shang{-}Wen Cheng}.}
  \bibinfo{year}{2009}\natexlab{}.
\newblock \showarticletitle{Software Architecture-Based Self-Adaptation}.
\newblock In \bibinfo{booktitle}{\emph{Autonomic Computing and Networking}},
  \bibfield{editor}{\bibinfo{person}{Yan Zhang},
  \bibinfo{person}{Laurence~Tianruo Yang}, {and} \bibinfo{person}{Mieso~K.
  Denko}} (Eds.). \bibinfo{publisher}{Springer}, \bibinfo{pages}{31--55}.
\newblock


\bibitem[\protect\citeauthoryear{Gay, Hartert, and Vissicchio}{Gay
  et~al\mbox{.}}{2017}]%
        {Gay:17}
\bibfield{author}{\bibinfo{person}{Steven Gay}, \bibinfo{person}{Renaud
  Hartert}, {and} \bibinfo{person}{Stefano Vissicchio}.}
  \bibinfo{year}{2017}\natexlab{}.
\newblock \showarticletitle{Expect the Unexpected: Sub-Second Optimization for
  Segment Routing}. In \bibinfo{booktitle}{\emph{Proceedings of the 2017 Annual
  IEEE International Conference on Computer Communications (INFOCOM'17)}}.
  \bibinfo{pages}{1--9}.
\newblock


\bibitem[\protect\citeauthoryear{Gheibi, Weyns, and Quin}{Gheibi
  et~al\mbox{.}}{2021}]%
        {DBLP:conf/seams/GheibiWQ21}
\bibfield{author}{\bibinfo{person}{Omid Gheibi}, \bibinfo{person}{Danny Weyns},
  {and} \bibinfo{person}{Federico Quin}.} \bibinfo{year}{2021}\natexlab{}.
\newblock \showarticletitle{On the Impact of Applying Machine Learning in the
  Decision-Making of Self-Adaptive Systems}. In \bibinfo{booktitle}{\emph{16th
  International Symposium on Software Engineering for Adaptive and
  Self-Managing Systems, {SEAMS}@ICSE 2021, Madrid, Spain, May 18-24, 2021}}.
  \bibinfo{publisher}{{IEEE}}, \bibinfo{pages}{104--110}.
\newblock


\bibitem[\protect\citeauthoryear{Ghobadi, Yeganeh, and Ganjali}{Ghobadi
  et~al\mbox{.}}{2012}]%
        {Yashar:12}
\bibfield{author}{\bibinfo{person}{Ma Ghobadi}, \bibinfo{person}{S.~Hassas
  Yeganeh}, {and} \bibinfo{person}{Y. Ganjali}.}
  \bibinfo{year}{2012}\natexlab{}.
\newblock \showarticletitle{Rethinking end-to-end congestion control in
  software-defined networks}. In \bibinfo{booktitle}{\emph{11th {ACM} Workshop
  on Hot Topics in Networks, HotNets-XI, Redmond, WA, {USA} - October 29 - 30,
  2012}}, \bibfield{editor}{\bibinfo{person}{Srikanth Kandula},
  \bibinfo{person}{Jitendra Padhye}, \bibinfo{person}{Emin~G{\"{u}}n Sirer},
  {and} \bibinfo{person}{Ramesh Govindan}} (Eds.). \bibinfo{publisher}{{ACM}},
  \bibinfo{pages}{61--66}.
\newblock


\bibitem[\protect\citeauthoryear{Guo, Wang, Yin, Shi, and Wu}{Guo
  et~al\mbox{.}}{2014}]%
        {6980429}
\bibfield{author}{\bibinfo{person}{Yingya Guo}, \bibinfo{person}{Zhiliang
  Wang}, \bibinfo{person}{Xia Yin}, \bibinfo{person}{Xingang Shi}, {and}
  \bibinfo{person}{Jianping Wu}.} \bibinfo{year}{2014}\natexlab{}.
\newblock \showarticletitle{Traffic Engineering in SDN/OSPF Hybrid Network}. In
  \bibinfo{booktitle}{\emph{2014 IEEE 22nd International Conference on Network
  Protocols}}. \bibinfo{pages}{563--568}.
\newblock


\bibitem[\protect\citeauthoryear{Haleplidis, Pentikousis, Denazis, Salim,
  Meyer, and Koufopavlou}{Haleplidis et~al\mbox{.}}{2015}]%
        {SDN:15}
\bibfield{author}{\bibinfo{person}{Evangelos Haleplidis},
  \bibinfo{person}{Kostas Pentikousis}, \bibinfo{person}{Spyros~G. Denazis},
  \bibinfo{person}{Jamal~Hadi Salim}, \bibinfo{person}{David Meyer}, {and}
  \bibinfo{person}{Odysseas~G. Koufopavlou}.} \bibinfo{year}{2015}\natexlab{}.
\newblock \bibinfo{booktitle}{\emph{Software-Defined Networking ({SDN}): Layers
  and Architecture Terminology}}.
\newblock \bibinfo{type}{Information} RFC 7426. \bibinfo{institution}{Internet
  Research Task Force (IRTF)}.
\newblock


\bibitem[\protect\citeauthoryear{Han, Dong, Guo, Shu, and Wu}{Han
  et~al\mbox{.}}{2015}]%
        {DBLP:journals/wicomm/HanDGSW15}
\bibfield{author}{\bibinfo{person}{Guangjie Han}, \bibinfo{person}{Yuhui Dong},
  \bibinfo{person}{Hui Guo}, \bibinfo{person}{Lei Shu}, {and}
  \bibinfo{person}{Dapeng Wu}.} \bibinfo{year}{2015}\natexlab{}.
\newblock \showarticletitle{Cross-layer optimized routing in wireless sensor
  networks with duty cycle and energy harvesting}.
\newblock \bibinfo{journal}{\emph{Wirel. Commun. Mob. Comput.}}
  \bibinfo{volume}{15}, \bibinfo{number}{16} (\bibinfo{year}{2015}),
  \bibinfo{pages}{1957--1981}.
\newblock


\bibitem[\protect\citeauthoryear{Harman, Burke, Clark, and Yao}{Harman
  et~al\mbox{.}}{2012}]%
        {Harman:12}
\bibfield{author}{\bibinfo{person}{Mark Harman}, \bibinfo{person}{Edmund
  Burke}, \bibinfo{person}{John Clark}, {and} \bibinfo{person}{Xin Yao}.}
  \bibinfo{year}{2012}\natexlab{}.
\newblock \showarticletitle{Dynamic Adaptive Search Based Software
  Engineering}. In \bibinfo{booktitle}{\emph{Proceedings of the 2012 ACM-IEEE
  International Symposium on Empirical Software Engineering and Measurement
  (ESEM'12)}}. \bibinfo{pages}{1--8}.
\newblock


\bibitem[\protect\citeauthoryear{Harman, McMinn, Souza, and Yoo}{Harman
  et~al\mbox{.}}{2010}]%
        {Harman2010SearchBS}
\bibfield{author}{\bibinfo{person}{M. Harman}, \bibinfo{person}{P. McMinn},
  \bibinfo{person}{J. Souza}, {and} \bibinfo{person}{S. Yoo}.}
  \bibinfo{year}{2010}\natexlab{}.
\newblock \showarticletitle{Search Based Software Engineering: Techniques,
  Taxonomy, Tutorial}. In \bibinfo{booktitle}{\emph{LASER Summer School}}.
\newblock


\bibitem[\protect\citeauthoryear{He, Rozner, Agarwal, Gu, Felter, Carter, and
  Akella}{He et~al\mbox{.}}{2016}]%
        {He:16}
\bibfield{author}{\bibinfo{person}{Keqiang He}, \bibinfo{person}{Eric Rozner},
  \bibinfo{person}{Kanak Agarwal}, \bibinfo{person}{Yu~(Jason) Gu},
  \bibinfo{person}{Wes Felter}, \bibinfo{person}{John~B. Carter}, {and}
  \bibinfo{person}{Aditya Akella}.} \bibinfo{year}{2016}\natexlab{}.
\newblock \showarticletitle{{AC/DC} {TCP}: Virtual Congestion Control
  Enforcement for Datacenter Networks}. In
  \bibinfo{booktitle}{\emph{Proceedings of the 2016 ACM Conference on Special
  Interest Group on Data Communication (SIGCOMM'16)}}.
  \bibinfo{pages}{244--257}.
\newblock


\bibitem[\protect\citeauthoryear{Hong, Kandula, Mahajan, Zhang, Gill, Nanduri,
  and Wattenhofer}{Hong et~al\mbox{.}}{2013}]%
        {Hong:13}
\bibfield{author}{\bibinfo{person}{Chi{-}Yao Hong}, \bibinfo{person}{Srikanth
  Kandula}, \bibinfo{person}{Ratul Mahajan}, \bibinfo{person}{Ming Zhang},
  \bibinfo{person}{Vijay Gill}, \bibinfo{person}{Mohan Nanduri}, {and}
  \bibinfo{person}{Roger Wattenhofer}.} \bibinfo{year}{2013}\natexlab{}.
\newblock \showarticletitle{Achieving High Utilization with Software-driven
  {WAN}}. In \bibinfo{booktitle}{\emph{Proceedings of the 2013 ACM Conference
  on Special Interest Group on Data Communication (SIGCOMM'13)}}.
  \bibinfo{pages}{15--26}.
\newblock


\bibitem[\protect\citeauthoryear{Huang, Liang, Xu, Xu, Guo, and Xu}{Huang
  et~al\mbox{.}}{2016}]%
        {Huang:16}
\bibfield{author}{\bibinfo{person}{Meitian Huang}, \bibinfo{person}{Weifa
  Liang}, \bibinfo{person}{Zichuan Xu}, \bibinfo{person}{Wenzheng Xu},
  \bibinfo{person}{Song Guo}, {and} \bibinfo{person}{Yinlong Xu}.}
  \bibinfo{year}{2016}\natexlab{}.
\newblock \showarticletitle{Dynamic Routing for Network Throughput Maximization
  in Software-Defined Networks}. In \bibinfo{booktitle}{\emph{Proceedings of
  the 2016 Annual IEEE International Conference on Computer Communications
  (INFOCOM'16)}}. \bibinfo{pages}{1--9}.
\newblock


\bibitem[\protect\citeauthoryear{Hung}{Hung}{2017}]%
        {IoT17}
\bibfield{author}{\bibinfo{person}{Mark Hung}.}
  \bibinfo{year}{2017}\natexlab{}.
\newblock \bibinfo{title}{Leading the IoT}.
\newblock \bibinfo{howpublished}{Documentation at
  \url{https://www.gartner.com/imagesrv/books/iot/iotEbook_digital.pdf}}.
\newblock


\bibitem[\protect\citeauthoryear{Iftikhar, Ramachandran, Bollans{\'{e}}e,
  Weyns, and Hughes}{Iftikhar et~al\mbox{.}}{2017}]%
        {Iftikhar:17}
\bibfield{author}{\bibinfo{person}{Muhammad~Usman Iftikhar},
  \bibinfo{person}{Gowri~Sankar Ramachandran}, \bibinfo{person}{Pablo
  Bollans{\'{e}}e}, \bibinfo{person}{Danny Weyns}, {and} \bibinfo{person}{Danny
  Hughes}.} \bibinfo{year}{2017}\natexlab{}.
\newblock \showarticletitle{{DeltaIoT}: {A} Self-Adaptive Internet of Things
  Exemplar}. In \bibinfo{booktitle}{\emph{Proceedings of the 12th International
  Symposium on Software Engineering for Adaptive and Self-Managing Systems
  {SEAMS'17}}}. \bibinfo{pages}{76--82}.
\newblock


\bibitem[\protect\citeauthoryear{Jahan, Riley, Walter, Gamble, Pasco, McKinley,
  and Cheng}{Jahan et~al\mbox{.}}{2020}]%
        {JahanRWGPMC20}
\bibfield{author}{\bibinfo{person}{Sharmin Jahan}, \bibinfo{person}{Ian Riley},
  \bibinfo{person}{Charles Walter}, \bibinfo{person}{Rose~F. Gamble},
  \bibinfo{person}{Matt Pasco}, \bibinfo{person}{Philip~K. McKinley}, {and}
  \bibinfo{person}{Betty H.~C. Cheng}.} \bibinfo{year}{2020}\natexlab{}.
\newblock \showarticletitle{{MAPE-K/MAPE-SAC:} An interaction framework for
  adaptive systems with security assurance cases}.
\newblock \bibinfo{journal}{\emph{Future Gener. Comput. Syst.}}
  \bibinfo{volume}{109} (\bibinfo{year}{2020}), \bibinfo{pages}{197--209}.
\newblock


\bibitem[\protect\citeauthoryear{Kephart and Chess}{Kephart and Chess}{2003}]%
        {kephart:03}
\bibfield{author}{\bibinfo{person}{Jeffrey~O. Kephart} {and}
  \bibinfo{person}{David~M. Chess}.} \bibinfo{year}{2003}\natexlab{}.
\newblock \showarticletitle{The Vision of Autonomic Computing}.
\newblock \bibinfo{journal}{\emph{Computer}} \bibinfo{volume}{36},
  \bibinfo{number}{1} (\bibinfo{year}{2003}), \bibinfo{pages}{41--50}.
\newblock


\bibitem[\protect\citeauthoryear{Koza and Koza}{Koza and Koza}{1992}]%
        {koza1992genetic}
\bibfield{author}{\bibinfo{person}{John~R Koza} {and} \bibinfo{person}{John~R
  Koza}.} \bibinfo{year}{1992}\natexlab{}.
\newblock \bibinfo{booktitle}{\emph{Genetic programming: on the programming of
  computers by means of natural selection}}. Vol.~\bibinfo{volume}{1}.
\newblock \bibinfo{publisher}{MIT press}.
\newblock


\bibitem[\protect\citeauthoryear{Krupitzer, Breitbach, Roth, VanSyckel,
  Schiele, and Becker}{Krupitzer et~al\mbox{.}}{2018}]%
        {Krupitzer:18}
\bibfield{author}{\bibinfo{person}{Christian Krupitzer},
  \bibinfo{person}{Martin Breitbach}, \bibinfo{person}{Felix~Maximilian Roth},
  \bibinfo{person}{Sebastian VanSyckel}, \bibinfo{person}{Gregor Schiele},
  {and} \bibinfo{person}{Christian Becker}.} \bibinfo{year}{2018}\natexlab{}.
\newblock \bibinfo{booktitle}{\emph{A Survey on Engineering Approaches for
  Self-Adaptive Systems (Extended Version)}}.
\newblock \bibinfo{type}{{T}echnical {R}eport}.
  \bibinfo{institution}{University of Mannheim}. \bibinfo{pages}{1--33} pages.
\newblock


\bibitem[\protect\citeauthoryear{Lantz, Heller, and McKeown}{Lantz
  et~al\mbox{.}}{2010}]%
        {Lantz:10}
\bibfield{author}{\bibinfo{person}{Bob Lantz}, \bibinfo{person}{Brandon
  Heller}, {and} \bibinfo{person}{Nick McKeown}.}
  \bibinfo{year}{2010}\natexlab{}.
\newblock \showarticletitle{A Network in a Laptop: Rapid Prototyping for
  Software-defined Networks}. In \bibinfo{booktitle}{\emph{Proceedings of the
  9th ACM SIGCOMM Workshop on Hot Topics in Networks (HotNets'10)}}.
  \bibinfo{pages}{19:1--19:6}.
\newblock


\bibitem[\protect\citeauthoryear{Lin, Teng, Hsu, Liao, and Lai}{Lin
  et~al\mbox{.}}{2016}]%
        {Lin:16}
\bibfield{author}{\bibinfo{person}{Ying{-}Dar Lin}, \bibinfo{person}{Hung{-}Yi
  Teng}, \bibinfo{person}{Chia{-}Rong Hsu}, \bibinfo{person}{Chun{-}Chieh
  Liao}, {and} \bibinfo{person}{Yuan{-}Cheng Lai}.}
  \bibinfo{year}{2016}\natexlab{}.
\newblock \showarticletitle{Fast Failover and Switchover for Link Failures and
  Congestion in Software Defined Networks}. In
  \bibinfo{booktitle}{\emph{Proceedings of the 2016 {IEEE} International
  Conference on Communications (ICC'16)}}. \bibinfo{pages}{1--6}.
\newblock


\bibitem[\protect\citeauthoryear{Liu, Zhou, Guo, and Qi}{Liu
  et~al\mbox{.}}{2018}]%
        {DBLP:conf/cscwd/LiuZGQ18}
\bibfield{author}{\bibinfo{person}{Lin Liu}, \bibinfo{person}{Jiantao Zhou},
  \bibinfo{person}{Xiaoyong Guo}, {and} \bibinfo{person}{Rui{-}dong Qi}.}
  \bibinfo{year}{2018}\natexlab{}.
\newblock \showarticletitle{A Method for Calculating Link Weight Dynamically by
  Entropy of Information in {SDN}}. In \bibinfo{booktitle}{\emph{22nd {IEEE}
  International Conference on Computer Supported Cooperative Work in Design,
  {CSCWD} 2018, Nanjing, China, May 9-11, 2018}}. \bibinfo{publisher}{{IEEE}},
  \bibinfo{pages}{535--540}.
\newblock


\bibitem[\protect\citeauthoryear{Lopes, Santos, Fidalgo, and Fernandes}{Lopes
  et~al\mbox{.}}{2016}]%
        {Lopes:16}
\bibfield{author}{\bibinfo{person}{Felipe~A. Lopes}, \bibinfo{person}{Marcelo
  Santos}, \bibinfo{person}{Robson Fidalgo}, {and} \bibinfo{person}{Stenio
  F.~L. Fernandes}.} \bibinfo{year}{2016}\natexlab{}.
\newblock \showarticletitle{A Software Engineering Perspective on {SDN}
  Programmability}.
\newblock \bibinfo{journal}{\emph{{IEEE} Communications Surveys and Tutorials}}
  \bibinfo{volume}{18}, \bibinfo{number}{2} (\bibinfo{year}{2016}),
  \bibinfo{pages}{1255--1272}.
\newblock


\bibitem[\protect\citeauthoryear{Luke}{Luke}{2013}]%
        {Luke:13}
\bibfield{author}{\bibinfo{person}{Sean Luke}.}
  \bibinfo{year}{2013}\natexlab{}.
\newblock \bibinfo{booktitle}{\emph{Essentials of Metaheuristics}
  (\bibinfo{edition}{second} ed.)}.
\newblock \bibinfo{publisher}{Lulu}.
\newblock
\newblock
\shownote{Available for free at
  http://cs.gmu.edu/$\sim$sean/book/metaheuristics/.}


\bibitem[\protect\citeauthoryear{Luke and Panait}{Luke and Panait}{2006}]%
        {LukeP06}
\bibfield{author}{\bibinfo{person}{Sean Luke} {and} \bibinfo{person}{Liviu
  Panait}.} \bibinfo{year}{2006}\natexlab{}.
\newblock \showarticletitle{A Comparison of Bloat Control Methods for Genetic
  Programming}.
\newblock \bibinfo{journal}{\emph{Evol. Comput.}} \bibinfo{volume}{14},
  \bibinfo{number}{3} (\bibinfo{year}{2006}), \bibinfo{pages}{309--344}.
\newblock


\bibitem[\protect\citeauthoryear{Mao, Fadlullah, Tang, Kato, Akashi, Inoue, and
  Mizutani}{Mao et~al\mbox{.}}{2017}]%
        {DBLP:conf/globecom/MaoFTKAIM17}
\bibfield{author}{\bibinfo{person}{Bomin Mao}, \bibinfo{person}{Zubair~Md.
  Fadlullah}, \bibinfo{person}{Fengxiao Tang}, \bibinfo{person}{Nei Kato},
  \bibinfo{person}{Osamu Akashi}, \bibinfo{person}{Takeru Inoue}, {and}
  \bibinfo{person}{Kimihiro Mizutani}.} \bibinfo{year}{2017}\natexlab{}.
\newblock \showarticletitle{A Tensor Based Deep Learning Technique for
  Intelligent Packet Routing}. In \bibinfo{booktitle}{\emph{2017 {IEEE} Global
  Communications Conference, {GLOBECOM} 2017, Singapore, December 4-8, 2017}}.
  \bibinfo{publisher}{{IEEE}}, \bibinfo{pages}{1--6}.
\newblock


\bibitem[\protect\citeauthoryear{Mathis and Mahdavi}{Mathis and
  Mahdavi}{1996}]%
        {Mathis:96}
\bibfield{author}{\bibinfo{person}{Matthew Mathis} {and}
  \bibinfo{person}{Jamshid Mahdavi}.} \bibinfo{year}{1996}\natexlab{}.
\newblock \showarticletitle{Forward Acknowledgement: Refining {TCP} Congestion
  Control}. In \bibinfo{booktitle}{\emph{Proceedings of the 1996 ACM Conference
  on Special Interest Group on Data Communication (SIGCOMM'96)}}.
  \bibinfo{pages}{281--291}.
\newblock


\bibitem[\protect\citeauthoryear{Matinnejad, Nejati, Briand, and
  Bruckmann}{Matinnejad et~al\mbox{.}}{2015}]%
        {paper4}
\bibfield{author}{\bibinfo{person}{Reza Matinnejad}, \bibinfo{person}{Shiva
  Nejati}, \bibinfo{person}{Lionel~C. Briand}, {and} \bibinfo{person}{Thomas
  Bruckmann}.} \bibinfo{year}{2015}\natexlab{}.
\newblock \showarticletitle{Effective test suites for mixed discrete-continuous
  stateflow controllers}. In \bibinfo{booktitle}{\emph{Proceedings of the 2015
  10th Joint Meeting on Foundations of Software Engineering, {ESEC/FSE} 2015}}.
  \bibinfo{publisher}{{ACM}}, \bibinfo{pages}{84--95}.
\newblock


\bibitem[\protect\citeauthoryear{Menasc\'{e}, Gomaa, Malek, and {a}o
  P.~Sousaa}{Menasc\'{e} et~al\mbox{.}}{2011}]%
        {Menasce:11}
\bibfield{author}{\bibinfo{person}{Daniel~A. Menasc\'{e}},
  \bibinfo{person}{Hassan Gomaa}, \bibinfo{person}{Sam Malek}, {and}
  \bibinfo{person}{Jo\ {a}o P.~Sousaa}.} \bibinfo{year}{2011}\natexlab{}.
\newblock \showarticletitle{{SASSY}: {A} Framework for Self-Architecting
  Service-Oriented Systems}.
\newblock \bibinfo{journal}{\emph{IEEE Software}} \bibinfo{volume}{28},
  \bibinfo{number}{6} (\bibinfo{year}{2011}), \bibinfo{pages}{78--85}.
\newblock


\bibitem[\protect\citeauthoryear{Menghi, Nejati, Briand, and Parache}{Menghi
  et~al\mbox{.}}{2020}]%
        {paper3}
\bibfield{author}{\bibinfo{person}{Claudio Menghi}, \bibinfo{person}{Shiva
  Nejati}, \bibinfo{person}{Lionel~C. Briand}, {and}
  \bibinfo{person}{Yago~Isasi Parache}.} \bibinfo{year}{2020}\natexlab{}.
\newblock \showarticletitle{Approximation-refinement testing of
  compute-intensive cyber-physical models: an approach based on system
  identification}. In \bibinfo{booktitle}{\emph{{ICSE} '20: 42nd International
  Conference on Software Engineering}}. \bibinfo{publisher}{{ACM}},
  \bibinfo{pages}{372--384}.
\newblock


\bibitem[\protect\citeauthoryear{Menghi, Nejati, Gaaloul, and Briand}{Menghi
  et~al\mbox{.}}{2019}]%
        {paper5}
\bibfield{author}{\bibinfo{person}{Claudio Menghi}, \bibinfo{person}{Shiva
  Nejati}, \bibinfo{person}{Khouloud Gaaloul}, {and} \bibinfo{person}{Lionel~C.
  Briand}.} \bibinfo{year}{2019}\natexlab{}.
\newblock \showarticletitle{Generating automated and online test oracles for
  Simulink models with continuous and uncertain behaviors}. In
  \bibinfo{booktitle}{\emph{Proceedings of the {ACM} Joint Meeting on European
  Software Engineering Conference and Symposium on the Foundations of Software
  Engineering, {ESEC/SIGSOFT} {FSE} 2019}}. \bibinfo{publisher}{{ACM}},
  \bibinfo{pages}{27--38}.
\newblock


\bibitem[\protect\citeauthoryear{Moreno, C{\'{a}}mara, Garlan, and
  Schmerl}{Moreno et~al\mbox{.}}{2015}]%
        {Moreno:15}
\bibfield{author}{\bibinfo{person}{Gabriel~A. Moreno}, \bibinfo{person}{Javier
  C{\'{a}}mara}, \bibinfo{person}{David Garlan}, {and}
  \bibinfo{person}{Bradley~R. Schmerl}.} \bibinfo{year}{2015}\natexlab{}.
\newblock \showarticletitle{Proactive self-adaptation under uncertainty: a
  probabilistic model checking approach}. In
  \bibinfo{booktitle}{\emph{Proceedings of the 2015 10th Joint Meeting on
  Foundations of Software Engineering, {ESEC/FSE} 2015, Bergamo, Italy, August
  30 - September 4, 2015}}, \bibfield{editor}{\bibinfo{person}{Elisabetta~Di
  Nitto}, \bibinfo{person}{Mark Harman}, {and} \bibinfo{person}{Patrick
  Heymans}} (Eds.). \bibinfo{publisher}{{ACM}}, \bibinfo{pages}{1--12}.
\newblock


\bibitem[\protect\citeauthoryear{Nejati}{Nejati}{2021}]%
        {paper1}
\bibfield{author}{\bibinfo{person}{Shiva Nejati}.}
  \bibinfo{year}{2021}\natexlab{}.
\newblock \showarticletitle{Next-Generation Software Verification: An {AI}
  Perspective}.
\newblock \bibinfo{journal}{\emph{{IEEE} Software}} \bibinfo{volume}{38},
  \bibinfo{number}{3} (\bibinfo{year}{2021}), \bibinfo{pages}{126--130}.
\newblock


\bibitem[\protect\citeauthoryear{Noormohammadpour and
  Raghavendra}{Noormohammadpour and Raghavendra}{2018}]%
        {Noormohammadpour18}
\bibfield{author}{\bibinfo{person}{Mohammad Noormohammadpour} {and}
  \bibinfo{person}{Cauligi~S. Raghavendra}.} \bibinfo{year}{2018}\natexlab{}.
\newblock \showarticletitle{Datacenter Traffic Control: Understanding
  Techniques and Tradeoffs}.
\newblock \bibinfo{journal}{\emph{{IEEE} Commun. Surv. Tutorials}}
  \bibinfo{volume}{20}, \bibinfo{number}{2} (\bibinfo{year}{2018}),
  \bibinfo{pages}{1492--1525}.
\newblock


\bibitem[\protect\citeauthoryear{Paucar and Bencomo}{Paucar and
  Bencomo}{2019}]%
        {PaucarB19}
\bibfield{author}{\bibinfo{person}{Luis Hern{\'{a}}n~Garc{\'{\i}}a Paucar}
  {and} \bibinfo{person}{Nelly Bencomo}.} \bibinfo{year}{2019}\natexlab{}.
\newblock \showarticletitle{Knowledge Base {K} Models to Support Trade-Offs for
  Self-Adaptation using Markov Processes}. In \bibinfo{booktitle}{\emph{13th
  {IEEE} International Conference on Self-Adaptive and Self-Organizing Systems,
  {SASO} 2019, Umea, Sweden, June 16-20, 2019}}. \bibinfo{publisher}{{IEEE}},
  \bibinfo{pages}{11--16}.
\newblock


\bibitem[\protect\citeauthoryear{Poli and B.~Langdon}{Poli and
  B.~Langdon}{1998}]%
        {poli1998schema}
\bibfield{author}{\bibinfo{person}{Riccardo Poli} {and}
  \bibinfo{person}{William B.~Langdon}.} \bibinfo{year}{1998}\natexlab{}.
\newblock \showarticletitle{Schema Theory for Genetic Programming with
  One-Point Crossover and Point Mutation}.
\newblock In \bibinfo{booktitle}{\emph{Evolutionary Computation}}.
  Vol.~\bibinfo{volume}{6}. \bibinfo{pages}{231--252}.
\newblock


\bibitem[\protect\citeauthoryear{Poli and Langdon}{Poli and Langdon}{1998}]%
        {onepointcrossover}
\bibfield{author}{\bibinfo{person}{Riccardo Poli} {and}
  \bibinfo{person}{William~B Langdon}.} \bibinfo{year}{1998}\natexlab{}.
\newblock \showarticletitle{Genetic programming with one-point crossover}. In
  \bibinfo{booktitle}{\emph{Soft Computing in Engineering Design and
  Manufacturing}}. \bibinfo{publisher}{Springer}, \bibinfo{pages}{180--189}.
\newblock


\bibitem[\protect\citeauthoryear{Poli, Langdon, McPhee, and Koza}{Poli
  et~al\mbox{.}}{2008}]%
        {poli2008field}
\bibfield{author}{\bibinfo{person}{Riccardo Poli}, \bibinfo{person}{William~B
  Langdon}, \bibinfo{person}{Nicholas~F McPhee}, {and} \bibinfo{person}{John~R
  Koza}.} \bibinfo{year}{2008}\natexlab{}.
\newblock \bibinfo{booktitle}{\emph{A field guide to genetic programming}}.
\newblock \bibinfo{publisher}{Lulu. com}.
\newblock


\bibitem[\protect\citeauthoryear{Poularakis, Iosifidis, Smaragdakis, and
  Tassiulas}{Poularakis et~al\mbox{.}}{2019}]%
        {Poularakis:19}
\bibfield{author}{\bibinfo{person}{Konstantinos Poularakis},
  \bibinfo{person}{George Iosifidis}, \bibinfo{person}{Georgios Smaragdakis},
  {and} \bibinfo{person}{Leandros Tassiulas}.} \bibinfo{year}{2019}\natexlab{}.
\newblock \showarticletitle{Optimizing Gradual {SDN} Upgrades in {ISP}
  Networks}.
\newblock \bibinfo{journal}{\emph{{IEEE/ACM} Transactions on Networking}}
  \bibinfo{volume}{27}, \bibinfo{number}{1} (\bibinfo{year}{2019}),
  \bibinfo{pages}{288--301}.
\newblock


\bibitem[\protect\citeauthoryear{Priyadarsini, Mukherjee, Bera, Kumar, Jakaria,
  and Rahman}{Priyadarsini et~al\mbox{.}}{2019}]%
        {PriyadarsiniMBK19}
\bibfield{author}{\bibinfo{person}{M. Priyadarsini}, \bibinfo{person}{J.~C.
  Mukherjee}, \bibinfo{person}{P. Bera}, \bibinfo{person}{S. Kumar},
  \bibinfo{person}{A.~H.~M. Jakaria}, {and} \bibinfo{person}{M.~Ashiqur
  Rahman}.} \bibinfo{year}{2019}\natexlab{}.
\newblock \showarticletitle{An adaptive load balancing scheme for
  software-defined network controllers}.
\newblock \bibinfo{journal}{\emph{Comput. Networks}}  \bibinfo{volume}{164}
  (\bibinfo{year}{2019}).
\newblock


\bibitem[\protect\citeauthoryear{Quin, Weyns, Bamelis, Buttar, and
  Michiels}{Quin et~al\mbox{.}}{2019}]%
        {DBLP:conf/icse/QuinWBBM19}
\bibfield{author}{\bibinfo{person}{Federico Quin}, \bibinfo{person}{Danny
  Weyns}, \bibinfo{person}{Thomas Bamelis}, \bibinfo{person}{Sarpreet~Singh
  Buttar}, {and} \bibinfo{person}{Sam Michiels}.}
  \bibinfo{year}{2019}\natexlab{}.
\newblock \showarticletitle{Efficient analysis of large adaptation spaces in
  self-adaptive systems using machine learning}. In
  \bibinfo{booktitle}{\emph{Proceedings of the 14th International Symposium on
  Software Engineering for Adaptive and Self-Managing Systems, {SEAMS}@ICSE
  2019, Montreal, QC, Canada, May 25-31, 2019}}. \bibinfo{publisher}{{ACM}},
  \bibinfo{pages}{1--12}.
\newblock


\bibitem[\protect\citeauthoryear{Ramirez and Cheng}{Ramirez and Cheng}{2010}]%
        {Ramirez:10}
\bibfield{author}{\bibinfo{person}{Andres~J. Ramirez} {and}
  \bibinfo{person}{Betty H.~C. Cheng}.} \bibinfo{year}{2010}\natexlab{}.
\newblock \showarticletitle{Design Patterns for Developing Dynamically Adaptive
  Systems}. In \bibinfo{booktitle}{\emph{Proceedings of the 2010 Workshop on
  Software Engineering for Adaptive and Self-Managing Systems {SEAMS'10}}}.
  \bibinfo{pages}{49--58}.
\newblock


\bibitem[\protect\citeauthoryear{Ramirez, Knoester, Cheng, and
  McKinley}{Ramirez et~al\mbox{.}}{2009}]%
        {Ramirez:09}
\bibfield{author}{\bibinfo{person}{Andres~J. Ramirez},
  \bibinfo{person}{David~B. Knoester}, \bibinfo{person}{Betty~H.C. Cheng},
  {and} \bibinfo{person}{Philip~K. McKinley}.} \bibinfo{year}{2009}\natexlab{}.
\newblock \showarticletitle{Applying Genetic Algorithms to Decision Making in
  Autonomic Computing Systems}. In \bibinfo{booktitle}{\emph{Proceedings of the
  6th International Conference on Autonomic Computing (ICAC'09)}}.
  \bibinfo{pages}{97--106}.
\newblock


\bibitem[\protect\citeauthoryear{Rego, Sendra, Jim{\'{e}}nez, and Lloret}{Rego
  et~al\mbox{.}}{2017}]%
        {Rego:17}
\bibfield{author}{\bibinfo{person}{Albert Rego}, \bibinfo{person}{Sandra
  Sendra}, \bibinfo{person}{Jos{\'{e}}~Miguel Jim{\'{e}}nez}, {and}
  \bibinfo{person}{Jaime Lloret}.} \bibinfo{year}{2017}\natexlab{}.
\newblock \showarticletitle{{OSPF} routing protocol performance in Software
  Defined Networks}. In \bibinfo{booktitle}{\emph{Proceedings of the 2017 4th
  International Conference on Software Defined Systems (SDS'17)}}.
  \bibinfo{pages}{131--136}.
\newblock


\bibitem[\protect\citeauthoryear{R{\'{e}}tv{\'{a}}ri, N{\'{e}}meth, Chaparadza,
  and Szab{\'{o}}}{R{\'{e}}tv{\'{a}}ri et~al\mbox{.}}{2009}]%
        {RetvariNCS09}
\bibfield{author}{\bibinfo{person}{G. R{\'{e}}tv{\'{a}}ri}, \bibinfo{person}{F.
  N{\'{e}}meth}, \bibinfo{person}{R. Chaparadza}, {and} \bibinfo{person}{R.
  Szab{\'{o}}}.} \bibinfo{year}{2009}\natexlab{}.
\newblock \showarticletitle{{OSPF} for Implementing Self-adaptive Routing in
  Autonomic Networks: {A} Case Study}. In \bibinfo{booktitle}{\emph{Modelling
  Autonomic Communications Environments, Fourth {IEEE} International Workshop,
  {MACE} 2009, Venice, Italy, October 26-27, 2009. Proceedings}},
  \bibfield{editor}{\bibinfo{person}{Yacine Ghamri{-}Doudane}} (Ed.),
  Vol.~\bibinfo{volume}{5844}. \bibinfo{publisher}{Springer},
  \bibinfo{pages}{72--85}.
\newblock


\bibitem[\protect\citeauthoryear{Salehie and Tahvildari}{Salehie and
  Tahvildari}{2009}]%
        {DBLP:journals/taas/SalehieT09}
\bibfield{author}{\bibinfo{person}{Mazeiar Salehie} {and}
  \bibinfo{person}{Ladan Tahvildari}.} \bibinfo{year}{2009}\natexlab{}.
\newblock \showarticletitle{Self-adaptive software: Landscape and research
  challenges}.
\newblock \bibinfo{journal}{\emph{{ACM} Trans. Auton. Adapt. Syst.}}
  \bibinfo{volume}{4}, \bibinfo{number}{2} (\bibinfo{year}{2009}),
  \bibinfo{pages}{14:1--14:42}.
\newblock


\bibitem[\protect\citeauthoryear{Scott and Luke}{Scott and Luke}{2019}]%
        {DBLP:conf/gecco/ScottL19}
\bibfield{author}{\bibinfo{person}{Eric~O. Scott} {and} \bibinfo{person}{Sean
  Luke}.} \bibinfo{year}{2019}\natexlab{}.
\newblock \showarticletitle{{ECJ} at 20: toward a general metaheuristics
  toolkit}. In \bibinfo{booktitle}{\emph{Proceedings of the Genetic and
  Evolutionary Computation Conference Companion, {GECCO} 2019, Prague, Czech
  Republic, July 13-17, 2019}}, \bibfield{editor}{\bibinfo{person}{Manuel
  L{\'{o}}pez{-}Ib{\'{a}}{\~{n}}ez}, \bibinfo{person}{Anne Auger}, {and}
  \bibinfo{person}{Thomas St{\"{u}}tzle}} (Eds.). \bibinfo{publisher}{{ACM}},
  \bibinfo{pages}{1391--1398}.
\newblock


\bibitem[\protect\citeauthoryear{Shin, Nejati, Sabetzadeh, Briand, Arora, and
  Zimmer}{Shin et~al\mbox{.}}{2020}]%
        {ShinNSB0Z20}
\bibfield{author}{\bibinfo{person}{Seung~Yeob Shin}, \bibinfo{person}{Shiva
  Nejati}, \bibinfo{person}{Mehrdad Sabetzadeh}, \bibinfo{person}{Lionel~C.
  Briand}, \bibinfo{person}{Chetan Arora}, {and} \bibinfo{person}{Frank
  Zimmer}.} \bibinfo{year}{2020}\natexlab{}.
\newblock \showarticletitle{Dynamic adaptation of software-defined networks for
  IoT systems: a search-based approach}. In \bibinfo{booktitle}{\emph{{SEAMS}
  '20: {IEEE/ACM} 15th International Symposium on Software Engineering for
  Adaptive and Self-Managing Systems, Seoul, Republic of Korea, 29 June - 3
  July, 2020}}, \bibfield{editor}{\bibinfo{person}{Shinichi Honiden},
  \bibinfo{person}{Elisabetta~Di Nitto}, {and} \bibinfo{person}{Radu
  Calinescu}} (Eds.). \bibinfo{publisher}{{ACM}}, \bibinfo{pages}{137--148}.
\newblock


\bibitem[\protect\citeauthoryear{Sotiropoulos, Waeselynck, Guiochet, and
  Ingrand}{Sotiropoulos et~al\mbox{.}}{2017}]%
        {SotiropoulosWGI17}
\bibfield{author}{\bibinfo{person}{Thierry Sotiropoulos},
  \bibinfo{person}{H{\'{e}}l{\`{e}}ne Waeselynck},
  \bibinfo{person}{J{\'{e}}r{\'{e}}mie Guiochet}, {and}
  \bibinfo{person}{F{\'{e}}lix Ingrand}.} \bibinfo{year}{2017}\natexlab{}.
\newblock \showarticletitle{Can Robot Navigation Bugs Be Found in Simulation?
  {An} Exploratory Study}. In \bibinfo{booktitle}{\emph{Proc. of the 2017
  {IEEE} International Conference on Software Quality, Reliability and
  Security}}. \bibinfo{pages}{150--159}.
\newblock


\bibitem[\protect\citeauthoryear{Stein, Fr\"{o}mmgen, Kluge, L\"{o}ffler,
  Sch\"{u}rr, Buchmann, and M\"{u}hlh\"{a}user}{Stein et~al\mbox{.}}{2016}]%
        {Stein:16}
\bibfield{author}{\bibinfo{person}{Michael Stein}, \bibinfo{person}{Alexander
  Fr\"{o}mmgen}, \bibinfo{person}{Roland Kluge}, \bibinfo{person}{Frank
  L\"{o}ffler}, \bibinfo{person}{Andy Sch\"{u}rr}, \bibinfo{person}{Alejandro
  Buchmann}, {and} \bibinfo{person}{Max M\"{u}hlh\"{a}user}.}
  \bibinfo{year}{2016}\natexlab{}.
\newblock \showarticletitle{{TARL}: Modeling Topology Adaptations for
  Networking Applications}. In \bibinfo{booktitle}{\emph{Proceedings of the
  11th International Symposium on Software Engineering for Adaptive and
  Self-Managing Systems {SEAMS'16}}}. \bibinfo{pages}{57--63}.
\newblock


\bibitem[\protect\citeauthoryear{Vargha and Delaney}{Vargha and
  Delaney}{2000}]%
        {vargha:00}
\bibfield{author}{\bibinfo{person}{Andr{\'a}s Vargha} {and}
  \bibinfo{person}{Harold~D. Delaney}.} \bibinfo{year}{2000}\natexlab{}.
\newblock \showarticletitle{A critique and improvement of the CL common
  language effect size statistics of McGraw and Wong}.
\newblock \bibinfo{journal}{\emph{Journal of Educational and Behavioral
  Statistics}} \bibinfo{volume}{25}, \bibinfo{number}{2}
  (\bibinfo{year}{2000}), \bibinfo{pages}{101--132}.
\newblock


\bibitem[\protect\citeauthoryear{Veenhuis}{Veenhuis}{2013}]%
        {Veenhuis13}
\bibfield{author}{\bibinfo{person}{Christian Veenhuis}.}
  \bibinfo{year}{2013}\natexlab{}.
\newblock \showarticletitle{Structure-Based Constants in Genetic Programming}.
  In \bibinfo{booktitle}{\emph{Progress in Artificial Intelligence - 16th
  Portuguese Conference on Artificial Intelligence, {EPIA} 2013, Angra do
  Hero{\'{\i}}smo, Azores, Portugal, September 9-12, 2013. Proceedings}}
  \emph{(\bibinfo{series}{Lecture Notes in Computer Science},
  Vol.~\bibinfo{volume}{8154})}, \bibfield{editor}{\bibinfo{person}{Lu{\'{\i}}s
  Correia}, \bibinfo{person}{Lu{\'{\i}}s~Paulo Reis}, {and}
  \bibinfo{person}{Jos{\'{e}} Cascalho}} (Eds.). \bibinfo{publisher}{Springer},
  \bibinfo{pages}{126--137}.
\newblock


\bibitem[\protect\citeauthoryear{Weyns, Iftikhar, Hughes, and Matthys}{Weyns
  et~al\mbox{.}}{2018}]%
        {Weyns:18}
\bibfield{author}{\bibinfo{person}{Danny Weyns}, \bibinfo{person}{M.~Usman
  Iftikhar}, \bibinfo{person}{Danny Hughes}, {and} \bibinfo{person}{Nelson
  Matthys}.} \bibinfo{year}{2018}\natexlab{}.
\newblock \showarticletitle{Applying Architecture-Based Adaptation to Automate
  the Management of Internet-of-Things}. In
  \bibinfo{booktitle}{\emph{Proceedings of the 12th European Conference on
  Software Architecture (ECSA'18)}}. \bibinfo{pages}{49--67}.
\newblock


\bibitem[\protect\citeauthoryear{Zhu, Karim, Sharif, Xu, Li, Du, and
  Guizani}{Zhu et~al\mbox{.}}{2021}]%
        {DBLP:journals/csur/ZhuKSXLDG21}
\bibfield{author}{\bibinfo{person}{Liehuang Zhu}, \bibinfo{person}{Md.~Monjurul
  Karim}, \bibinfo{person}{Kashif Sharif}, \bibinfo{person}{Chang Xu},
  \bibinfo{person}{Fan Li}, \bibinfo{person}{Xiaojiang Du}, {and}
  \bibinfo{person}{Mohsen Guizani}.} \bibinfo{year}{2021}\natexlab{}.
\newblock \showarticletitle{{SDN} Controllers: {A} Comprehensive Analysis and
  Performance Evaluation Study}.
\newblock \bibinfo{journal}{\emph{{ACM} Comput. Surv.}} \bibinfo{volume}{53},
  \bibinfo{number}{6} (\bibinfo{year}{2021}), \bibinfo{pages}{133:1--133:40}.
\newblock


\bibitem[\protect\citeauthoryear{Zoghi, Shtern, Litoiu, and Ghanbari}{Zoghi
  et~al\mbox{.}}{2016}]%
        {Zoghi:16}
\bibfield{author}{\bibinfo{person}{Parisa Zoghi}, \bibinfo{person}{Mark
  Shtern}, \bibinfo{person}{Marin Litoiu}, {and} \bibinfo{person}{Hamoun
  Ghanbari}.} \bibinfo{year}{2016}\natexlab{}.
\newblock \showarticletitle{Designing Adaptive Applications Deployed on Cloud
  Environments}.
\newblock \bibinfo{journal}{\emph{ACM Transactions on Autonomous and Adaptive
  Systems (TAAS)}} \bibinfo{volume}{10}, \bibinfo{number}{4}
  (\bibinfo{year}{2016}), \bibinfo{pages}{25:1--25:26}.
\newblock


\end{thebibliography}
